\documentclass[useAMS,usenatbib]{mn2e}

\usepackage{latexsym}
\usepackage{afterpage}
\usepackage{longtable}
\usepackage[dvips]{graphicx}
\usepackage{natbib}
\usepackage{psfrag}
\usepackage{rotating}
\usepackage{amssymb}

\newcommand\ion[2]{#1{\scshape{#2}}}

\def\ApJ{{\it Astroph. J.}}

\def\MNRAS{{\it Mon. Not. R. Astr. Soc.}}
\def\AandA{{\it Astron. \& Astrophys.}}

\title[Flux density variations of radio sources in M82]{Flux density variations of radio sources in M82 over the last 3 decades}

\author[Gendre, Fenech, Beswick, Muxlow \& Argo]{M. A. Gendre$^{1}$\thanks{E-mail:
    mgendre@jb.man.ac.uk}, D. M. Fenech$^{2}$, R. J. Beswick$^{1}$, T. W. B. Muxlow$^{1}$ \& M. K. Argo$^{1,3}$\\ 
$^{1}$Jodrell Bank Center for Astrophysics, The University of Manchester, Oxford Rd, Manchester M13 9PL, United Kingdom\\ 
$^{2}$Department of Physics and Astronomy, University College London, Gower Street, London WC1E 6BT, United Kingdom\\
$^{3}$Netherlands Institute for Radio Astronomy (ASTRON), Postbus 2, 7990 AA Dwingeloo, The Netherlands\\}

\begin{document}

\date{Accepted . Received ; in original form }

\pagerange{\pageref{firstpage}--\pageref{lastpage}} \pubyear{}

\maketitle

\label{firstpage}

\begin{abstract}
This paper presents the results of the 2009-2010 monitoring sessions of the starburst galaxy M82, obtained with the Multi-Element Radio-Linked Interferometer Network (MERLIN) at 5\,GHz and {\it e}-MERLIN at 6\,GHz. Combining several 5\,GHz MERLIN epochs to form a map with 33.0 $\mu$Jy/bm noise level, 52 discrete sources, mostly supernova remnants and \ion{H}{ii} regions, are identified. These include three objects which were not detected in the 2002 5\,GHz MERLIN monitoring session: supernova SN2008iz, the transient source 43.78+59.3, and a new supernova remnant shell. Flux density variations, in the long (1981 to 2010), medium (2002 to 2010) and short (2009 to 2010) term, are investigated. We find that flux densities of SNRs in M82 stay constant in most of the sample ($\sim$95\%). In addition, aside from SN2008iz and the well-known variable source 41.95+57.5, two sources display short and medium term variations over the period 2009-2010. These sources being among the most compact SNR in M82, these flux density variations could be due to changes in the circumstellar and interstellar medium in which the shocks travel.
\end{abstract}


\section{Introduction}

\begin{figure*}
  \centerline{
    \includegraphics[angle=270,scale=0.65]{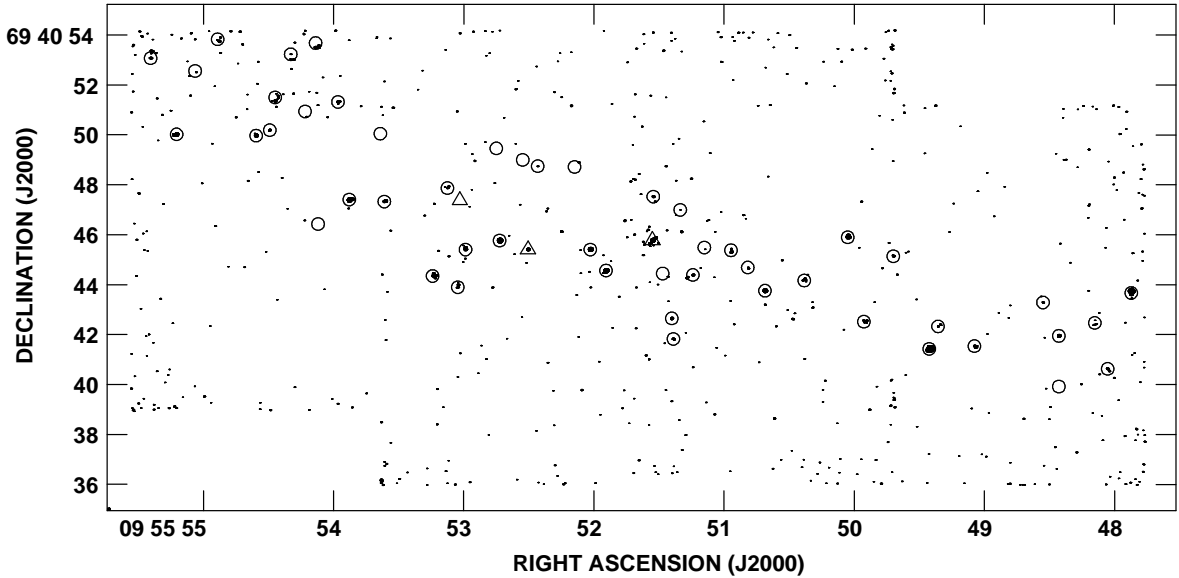}}
  \caption[]{\label{M82Panel}Position of the 52 sources in M82 identified from the 5\,GHz MERLIN combined data presented in this work (contour plot at $-$1, 1, 1.414, 2, 2.828, 4, 5.656, 8, 11.282, 16 $\times$ 99 $\mu$Jy/bm). Sources present in \cite{Fenech08} are represented by circles while new sources are represented by triangles.} 
\end{figure*}

M82 is one of the closest \citep[$d$ = 3.2 Mpc;][]{Burb64} starburst galaxies known, producing a large population of massive, rapidly evolving stars, and an equally large number of supernovae. The central regions of the majority of starbursts suffer from high visual extinction. Radio observations, which are unaffected by dust, are thus ideal for the detection and study of these supernovae or supernova remnants (SNRs).\\

Over the past 30 years, M82 has been subject to frequent radio monitoring at centimetre wavelengths with the Very Large Array \citep[VLA; e.g.][]{Kro00,All99} and the Multi-Element Radio-Linked Interferometer Network \citep[MERLIN; e.g.][]{Mux94,Fenech08}. With the detection of over 50 discrete objects, including over 30 SNRs, these regular observation programmes have the advantage of tracking the evolution of supernova remnants as their shells expand, which provides a way to investigate properties of the Inter-Stellar Medium \citep[see the review by][]{Wei02}. Since sources in M82 can be effectively treated as being at the same distance (their relative distances differing by only a few hundred parsecs, which is small compared with the overall distance of the galaxy), they can be studied with essentially the same linear resolution and luminosity sensitivity. Another advantage of studying extragalactic SNRs is that younger objects can be observed in M82 (from a few to a few hundred years) than in the Milky Way (until recently, the youngest was Cassiopeia A, which is over 300-yr old; however, \cite{Green08} discovered a younger galactic SNR $\sim$150-yr old). In addition, new sources such as SN2008iz \citep{Brun09} or the radio transient 43.78+59.3 \citep{Mux10} would not have been detected without regular radio monitoring. Finally, the number of SNRs detected provides a way of determining star formation rates in starburst galaxies, based on the number of sources in the galaxy, which is related to the supernova rate, and by assuming a given initial-mass function \citep[][]{Cond92,Bott12}.

Regular observation programmes also allow for the monitoring of flux density variability in sources \citep[e.g.][]{Kro00}, such as 41.95+57.5, which has shown a continued decrease in flux density since its first observation in 1965 \citep{Trot96,Mux05}.\\

This paper presents and discusses the results of the 2009-2010 monitoring sessions, which includes seven epochs at 5\,GHz with MERLIN and one epoch at 6\,GHz with {\it e}-MERLIN. These results have been supplemented with results from the literature \citep{Kro00,Fenech08} increasing the time-span of monitoring observations to 30 years. It is expected that most sources will show no (or only very little) flux density variations over the time periods presented here \citep{Ulv94,Wei02}. This work also intends to investigate the behaviour of SNR in M82 and compare it to previous models \citep[e.g.][]{Sea07}.

Details of the data reduction techniques are given in Section \ref{Proc}, while Section \ref{Monit} describes the source identification process and flux density measurements. Finally, flux density variations in the identified sources are discussed in Section \ref{Results}, including more detailed discussions about a few sources in Section \ref{PartSource}.

\section{MERLIN observations and image processing}\label{Proc}

\begin{table}
  \begin{center}
    \caption[]{\label{EpochTab}Details of 2009-2010 MERLIN (EP1-7), 2010 {\it e}-MERLIN (EP8) and archival MERLIN observations. ToS correspond to time on source (in hours)$^a$ and the rms noise is given in $\mu$Jy/bm$^b$. Unless indicated otherwise, the observation were done without the Lovell telescope.\\
      $^T$ Tabley Telescope not available.\\
      $^D$ Defford Telescope not available.\\
      $^L$ Including Lovell telescope.\\
     The table also displays, in the last column, measured values of flux density (in mJy)$^c$ of the phase calibrator 0955+697. The flux density for EP8, measured at 6.0\,GHz, was converted to 4.995\,GHz using $S \propto \nu^{\alpha}$ where $\alpha$=$-$0.23 \citep{Vol10} for this source.}
    \medskip
    \begin{tabular}{cccccc}
      \hline
      Ep. & Start date & \# & $^a$ToS & $^b$rms & $^c$$S_{int}$\\
            & of obs. & ant. &  &  & 0955+697 \\
      \hline                                                               
           & 2002       & 6               &            175.0& \phantom{1}23.0 & 70.0$\pm$0.2\\
           & 2005       & 7$^L$           &              9.0& \phantom{1}55.0 & 76.8$\pm$0.2\\
      &&&&&\\   
      EP1  & 2009-05-01 & 6\phantom{$^T$} & \phantom{1}65.6 & \phantom{1}61.0 & 79.4$\pm$0.2\\
      EP2  & 2009-05-26 & 5$^T$           & \phantom{1}12.2 &           150.0 & 80.7$\pm$0.4\\
      EP3  & 2009-07-28 & 5$^D$           &            110.2& \phantom{1}39.0 & 94.0$\pm$0.1\\
      EP4  & 2009-08-19 & 5$^D$           & \phantom{1}34.9 & \phantom{1}91.0 & 99.6$\pm$0.3\\
      EP5  & 2009-09-12 & 6\phantom{$^T$} & \phantom{1}28.9 & \phantom{1}70.0 & 92.7$\pm$0.3\\
      EP6  & 2009-09-25 & 6\phantom{$^T$} & \phantom{1}25.9 & \phantom{1}61.0 & 84.2$\pm$0.3\\
      EP7  & 2010-04-06 & 6\phantom{$^T$} & \phantom{10}8.8 &           100.0 & 99.3$\pm$0.4\\
      COMB.& 2009       &                 &           286.5 & \phantom{1}33.0 &        -    \\
      &&&&&\\                                                                  
      EP8  & 2010-12-17 & 6\phantom{$^T$} & \phantom{1}19.1 & \phantom{1}40.0 &125.2$\pm$0.4\\
      \hline
    \end{tabular}
  \end{center}
\end{table}
\begin{figure}
  \centerline{
    \includegraphics[scale=0.38]{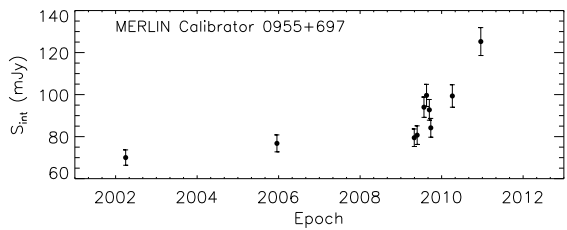}}
  \caption[]{\label{PHcalflux}Measured flux densities of the phase calibrator 0955+697 over the period 2002-2010.} 
\end{figure}

The 2009 monitoring campaign of M82 consisted of seven wide-field MERLIN \citep{Thomasson86} observations at a frequency of 4.994\,GHz, observed between May 2009 and April 2010. These were made using parallel hands of circular polarisation, and were correlated with a total bandwidth of 16\,MHz divided into 32 channels. Across all epochs combined a total on-source integration time of 286.5\,hr was used. Each observing epoch was reduced and analysed individually and a deep exposure map was produced by combining all of these data. In addition to these, a single {\it e}-MERLIN observation was included in this study. This observation was made in December 2010 as part of the {\it e}-MERLIN commissioning programme and used a total bandwidth of 512\,MHz with a median frequency of 6.26\,GHz. These data were correlated into four individual sub-bands each divided into 512 frequency channels. These e-MERLIN observations were observed prior the installation of the new {\it e}-MERLIN wide-band IF system in spring 2011 and consequently only one hand of polarisation was used and the data displayed reduced sensitivity in parts of the observing band. Further details of both the MERLIN and {\it e}-MERLIN observations, including archival data, are provided in Table~\ref{EpochTab}.

For both MERLIN and {\it e}-MERLIN data, the calibrator 3C286 was used to set the primary flux density scales and the point source calibrator OQ208 was used to determine the bandpass and relative gains of the antennas. Observations of the phase reference source, 0955+697, were interleaved throughout each of the observations and used to determine the telescope phases. Following phase calibration these data were re-weighted appropriately to account for the relative sensitivity of each MERLIN antenna. All epochs, as well as the combined 2009 epoch, were then imaged separately.

\begin{figure}
  \begin{minipage}{8cm}
  \centerline{
    \includegraphics[scale=0.4]{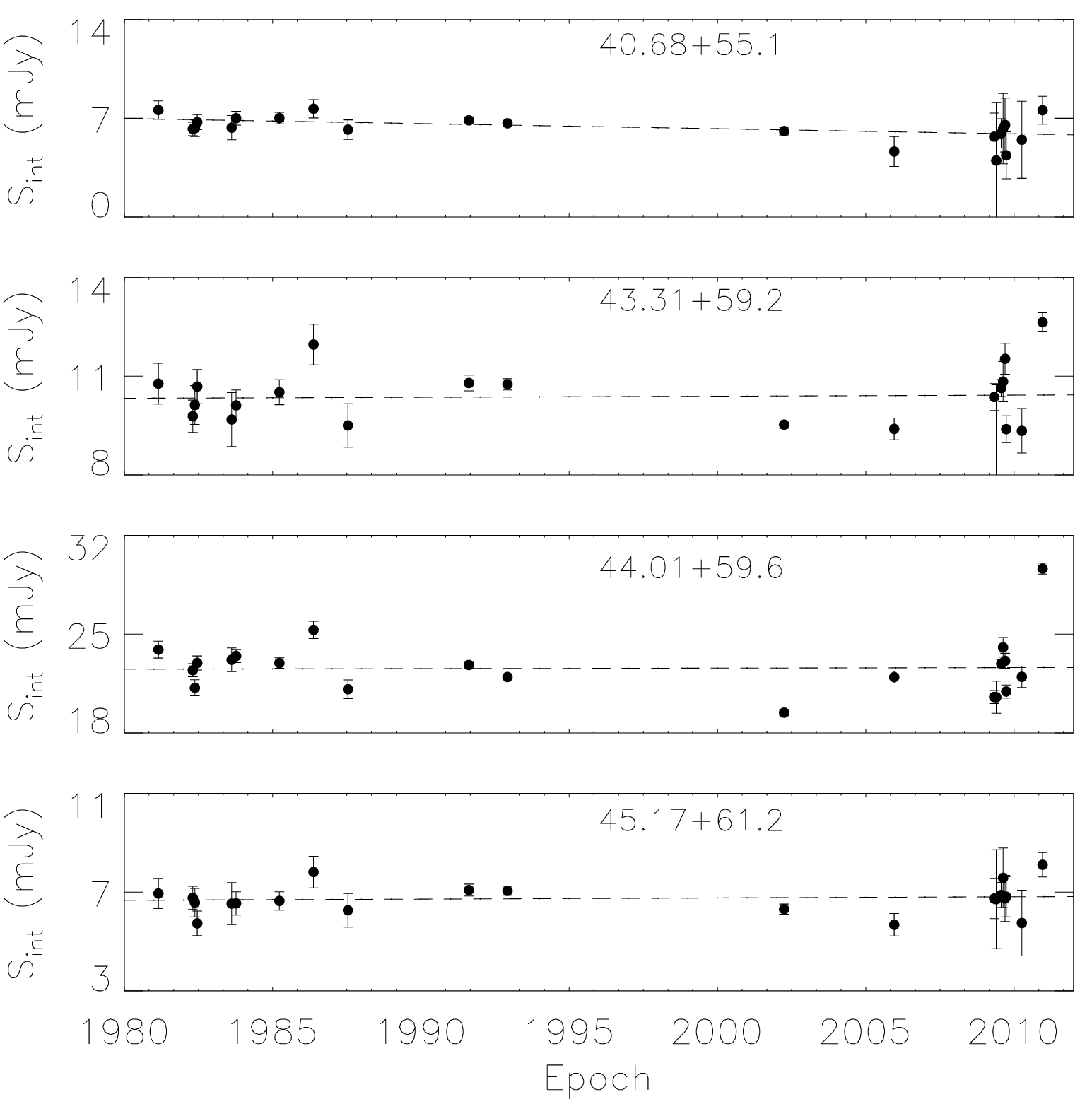}}
  \vspace{5mm}
  \end{minipage}
  \vfill
  \vspace{2mm}
  \begin{minipage}{8.cm}
  \centerline{
    \includegraphics[scale=0.43]{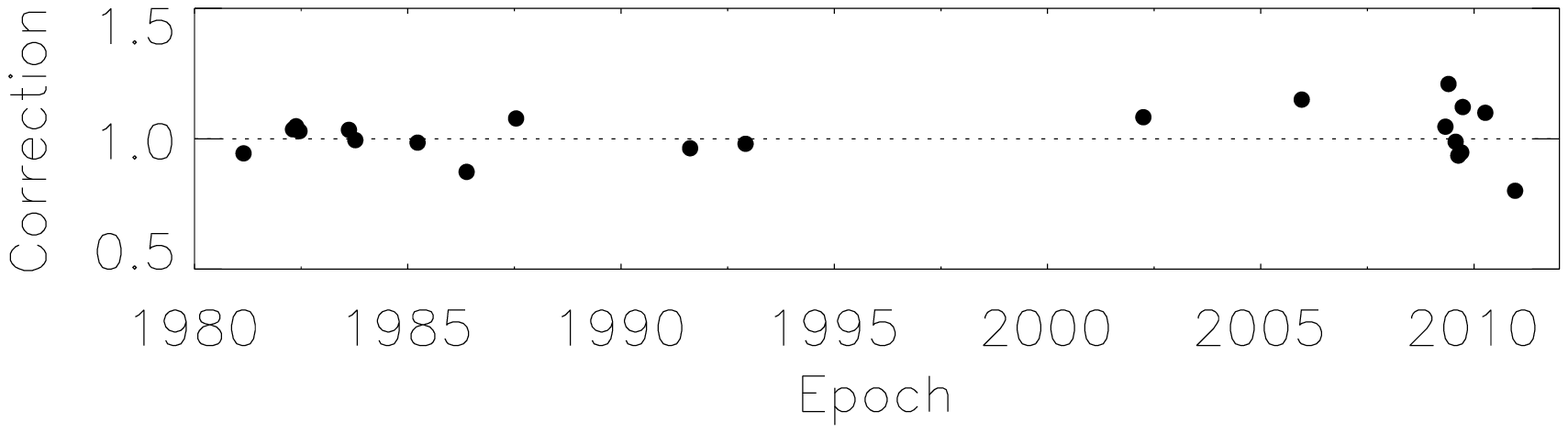}}
  \caption[]{\label{Norm}Top: Light curve for the bright stable sources (from top to bottom) 40.68+55.1, 43.31+59.2, 44.01+59.6 and 45.17+61.2 used for internal calibration, using data over the periods 1981-1991 \citep{Kro00} and 2002-2010 (present work). Bottom: Magnitude of the internal calibration corrections applied.}
  \end{minipage}
\end{figure}

\section{Source monitoring}\label{Monit}

\begin{table*}
  \begin{center}
    \caption[]{\label{FluxTab}Combined epoch (averaged over all 2009 data) normalised (as described in \S\ref{FMeas}) 2009 peak and integrated flux densities, for all 52 detected objects. Sources previously identified in the literature as SNRs or \ion{H}{ii} regions have been labelled accordingly. These identifications are taken from \cite{Mux94}, \cite{Wills97}, \cite{All99} and \cite{McD02}. The Right Ascension and Declination are offset from 09h55m00s and +69$^{\circ}$40$^{\prime}$00$^{\prime\prime}$(J2000), while the source name correspond to the equivalent offset in B1950 coordinates. Peak flux densities were measured with an error of $\pm$0.01 mJy. \\ $^{\ast}$SN2008iz \citep{Brun09} and transient source 43.78+59.3 \citep{Mux10}.\\ $^{\dagger}$Source absent from the 2002 data with no previous reference found.}
    \medskip
    \begin{tabular}{lcccccl}
      \hline
      \multicolumn{1}{c}{number} &\multicolumn{1}{c}{name} & \multicolumn{1}{c}{RA}      & \multicolumn{1}{c}{DEC}   & \multicolumn{1}{c}{Peak Flux} & \multicolumn{1}{c}{Integrated Flux} & \multicolumn{1}{c}{Comments}\\
                                 &                         &                             &                           & \multicolumn{1}{c}{Density}   & \multicolumn{1}{c}{Density}         & \multicolumn{1}{c}{}\\
                                 &                         & \multicolumn{2}{c}{(J2000)}                             & \multicolumn{1}{c}{(mJy/bm)}& \multicolumn{1}{c}{(mJy)} & \\
      \hline
      1  & \phantom{$^{\ast}$}39.10+57.3 & 47.87 & 43.7 &  \phantom{1}0.51  &   \phantom{1}4.50$\pm$0.65  &  SNR\\
      2  & \phantom{$^{\ast}$}39.28+54.1 & 48.05 & 40.6 &  \phantom{1}0.22  &   \phantom{1}0.23$\pm$0.05  &  \ion{H}{ii}\\
      3  & \phantom{$^{\ast}$}39.40+56.2 & 48.15 & 42.5 &  \phantom{1}0.22  &   \phantom{1}1.13$\pm$0.29  &  SNR\\
      4  & \phantom{$^{\ast}$}39.64+53.3 & 48.41 & 39.7 &  \phantom{1}0.16  &   \phantom{1}0.32$\pm$0.09  &  SNR\\
      5  & \phantom{$^{\ast}$}39.67+55.5 & 48.43 & 41.9 &  \phantom{1}0.34  &   \phantom{1}1.22$\pm$0.22  &  \ion{H}{ii}\\
      6  & \phantom{$^{\ast}$}39.77+56.9 & 48.55 & 43.3 &  \phantom{1}0.15  &   \phantom{1}0.36$\pm$0.10  &  SNR\\
      7  & \phantom{$^{\ast}$}40.32+55.2 & 49.07 & 41.5 &  \phantom{1}0.27  &   \phantom{1}1.35$\pm$0.27  &  SNR\\
      8  & \phantom{$^{\ast}$}40.61+56.3 & 49.33 & 42.3 &  \phantom{1}0.18  &   \phantom{1}1.27$\pm$0.33  &  SNR\\
      9  & \phantom{$^{\ast}$}40.68+55.1 & 49.42 & 41.4 &  \phantom{1}0.50  &   \phantom{1}7.19$\pm$0.90  &  SNR\\
      10 & \phantom{$^{\ast}$}40.94+58.8 & 49.69 & 45.2 &  \phantom{1}0.25  &   \phantom{1}0.73$\pm$0.17  &  \ion{H}{ii}\\
      11 & \phantom{$^{\ast}$}41.18+56.2 & 49.92 & 42.5 &  \phantom{1}0.23  &   \phantom{1}1.76$\pm$0.44  &  \ion{H}{ii}\\
      12 & \phantom{$^{\ast}$}41.30+59.6 & 50.04 & 45.9 &  \phantom{1}0.61  &   \phantom{1}2.54$\pm$0.29  &  SNR\\
      13 & \phantom{$^{\ast}$}41.64+57.9 & 50.38 & 44.2 &  \phantom{1}0.22  &   \phantom{1}1.02$\pm$0.22  &  \ion{H}{ii}\\
      14 & \phantom{$^{\ast}$}41.95+57.5 & 50.68 & 43.8 &            11.92  &             14.17$\pm$0.19  &  SNR(?)\\
      15 & \phantom{$^{\ast}$}42.08+58.4 & 50.82 & 44.7 &  \phantom{1}0.15  &   \phantom{1}0.34$\pm$0.09  &  \ion{H}{ii}\\
      16 & \phantom{$^{\ast}$}42.20+59.1 & 50.94 & 45.3 &  \phantom{1}0.32  &   \phantom{1}0.99$\pm$0.19  &  \ion{H}{ii}\\
      17 & \phantom{$^{\ast}$}42.43+59.5 & 51.15 & 45.6 &  \phantom{1}0.16  &   \phantom{1}0.38$\pm$0.15  &  Unknown\\
      18 & \phantom{$^{\ast}$}42.48+58.4 & 51.25 & 44.2 &  \phantom{1}0.31  &   \phantom{1}1.37$\pm$0.37  &  \ion{H}{ii}\\
      19 & \phantom{$^{\ast}$}42.61+60.7 & 51.35 & 46.9 &  \phantom{1}0.19  &   \phantom{1}0.67$\pm$0.16  &  SNR\\
      20 & \phantom{$^{\ast}$}42.67+55.6 & 51.37 & 41.8 &  \phantom{1}0.20  &   \phantom{1}0.81$\pm$0.19  &  SNR\\
      21 & \phantom{$^{\ast}$}42.67+56.3 & 51.40 & 42.6 &  \phantom{1}0.24  &   \phantom{1}0.63$\pm$0.13  &  SNR\\
      22 & \phantom{$^{\ast}$}42.69+58.2 & 51.44 & 44.3 &  \phantom{1}0.15  &   \phantom{1}0.21$\pm$0.06  &  \ion{H}{ii}\\
      23 & \phantom{$^{\ast}$}42.80+61.2 & 51.54 & 47.5 &  \phantom{1}0.23  &   \phantom{1}0.55$\pm$0.11  &  SNR\\
      24 &           $^{\ast}$42.81+59.6 & 51.55 & 45.8 &            31.60  &             33.57$\pm$0.17  &  SN\\
      25 & \phantom{$^{\ast}$}43.18+58.2 & 51.90 & 44.6 &  \phantom{1}1.24  &   \phantom{1}4.79$\pm$0.33  &  SNR\\
      26 & \phantom{$^{\ast}$}43.31+59.2 & 52.03 & 45.4 &  \phantom{1}6.60  &             12.52$\pm$0.22  &  SNR\\
      27 & \phantom{$^{\ast}$}43.40+62.6 & 52.14 & 48.7 &  \phantom{1}0.18  &   \phantom{1}0.61$\pm$0.16  &  SNR\\
      28 & \phantom{$^{\ast}$}43.72+62.6 & 52.43 & 48.8 &  \phantom{1}0.25  &   \phantom{1}0.42$\pm$0.09  &  SNR\\
      29 &           $^{\ast}$43.78+59.3 & 52.51 & 45.4 &  \phantom{1}0.80  &   \phantom{1}0.90$\pm$0.09  &  Trans.\\
      30 & \phantom{$^{\ast}$}43.82+62.8 & 52.54 & 49.0 &  \phantom{1}0.16  &   \phantom{1}0.33$\pm$0.09  &  SNR\\
      31 & \phantom{$^{\ast}$}44.01+59.6 & 52.72 & 45.8 &            13.07  &             25.88$\pm$0.25  &  SNR\\
      32 & \phantom{$^{\ast}$}44.08+63.1 & 52.75 & 49.5 &  \phantom{1}0.12  &   \phantom{1}0.15$\pm$0.06  &  SNR\\
      33 & \phantom{$^{\ast}$}44.28+59.3 & 52.99 & 45.5 &  \phantom{1}0.34  &   \phantom{1}1.72$\pm$0.32  &  SNR\\
      34 &         $^{\dagger}$44.28+61.1 & 53.00 & 47.4 & \phantom{1}0.17  &   \phantom{1}0.40$\pm$0.11  &  SNR(?)\\
      35 & \phantom{$^{\ast}$}44.34+57.8 & 53.05 & 43.9 &  \phantom{1}0.21  &   \phantom{1}0.98$\pm$0.23  &  SNR\\
      36 & \phantom{$^{\ast}$}44.40+61.8 & 53.13 & 47.9 &  \phantom{1}0.25  &   \phantom{1}1.19$\pm$0.25  &  SNR\\
      37 & \phantom{$^{\ast}$}44.51+58.2 & 53.23 & 44.3 &  \phantom{1}0.25  &   \phantom{1}2.38$\pm$0.54  &  SNR\\
      38 & \phantom{$^{\ast}$}44.89+61.2 & 53.61 & 47.3 &  \phantom{1}0.31  &   \phantom{1}1.16$\pm$0.21  &  SNR\\
      39 & \phantom{$^{\ast}$}44.93+64.0 & 53.64 & 50.0 &  \phantom{1}0.19  &   \phantom{1}0.39$\pm$0.10  &  \ion{H}{ii}\\
      40 & \phantom{$^{\ast}$}45.17+61.2 & 53.88 & 47.4 &  \phantom{1}3.11  &   \phantom{1}8.72$\pm$0.44  &  SNR\\
      41 & \phantom{$^{\ast}$}45.24+65.2 & 53.96 & 51.4 &  \phantom{1}0.21  &   \phantom{1}1.15$\pm$0.26  &  SNR\\
      42 & \phantom{$^{\ast}$}45.39+60.3 & 54.12 & 46.4 &  \phantom{1}0.17  &   \phantom{1}1.00$\pm$0.26  &  Unknown\\
      43 & \phantom{$^{\ast}$}45.42+67.4 & 54.13 & 53.5 &  \phantom{1}0.28  &   \phantom{1}2.56$\pm$0.55  &  SNR\\
      44 & \phantom{$^{\ast}$}45.52+64.7 & 54.22 & 50.9 &  \phantom{1}0.09  &   \phantom{1}0.12$\pm$0.05  &  SNR\\
      45 & \phantom{$^{\ast}$}45.62+67.0 & 54.32 & 53.1 &  \phantom{1}0.24  &   \phantom{1}1.54$\pm$0.37  &  \ion{H}{ii}\\
      46 & \phantom{$^{\ast}$}45.75+65.3 & 54.46 & 51.5 &  \phantom{1}0.20  &   \phantom{1}2.24$\pm$0.54  &  SNR\\
      47 & \phantom{$^{\ast}$}45.79+64.0 & 54.50 & 50.2 &  \phantom{1}0.21  &   \phantom{1}0.49$\pm$0.11  &  SNR\\
      48 & \phantom{$^{\ast}$}45.89+63.8 & 54.60 & 50.0 &  \phantom{1}0.58  &   \phantom{1}1.60$\pm$0.20  &  SNR\\
      49 & \phantom{$^{\ast}$}46.17+67.6 & 54.89 & 53.8 &  \phantom{1}0.24  &   \phantom{1}2.35$\pm$0.57  &  \ion{H}{ii}\\
      50 & \phantom{$^{\ast}$}46.34+66.2 & 55.06 & 52.5 &  \phantom{1}0.18  &   \phantom{1}0.67$\pm$0.17  &  Unknown\\
      51 & \phantom{$^{\ast}$}46.52+63.9 & 55.20 & 50.0 &  \phantom{1}0.42  &   \phantom{1}2.68$\pm$0.47  &  SNR\\
      52 & \phantom{$^{\ast}$}46.70+67.1 & 55.38 & 53.2 &  \phantom{1}0.22  &   \phantom{1}2.29$\pm$0.54  &  SNR\\
      \hline
    \end{tabular}
  \end{center}
\end{table*}
\begin{figure*}
    \centerline{
      \includegraphics[scale=0.88]{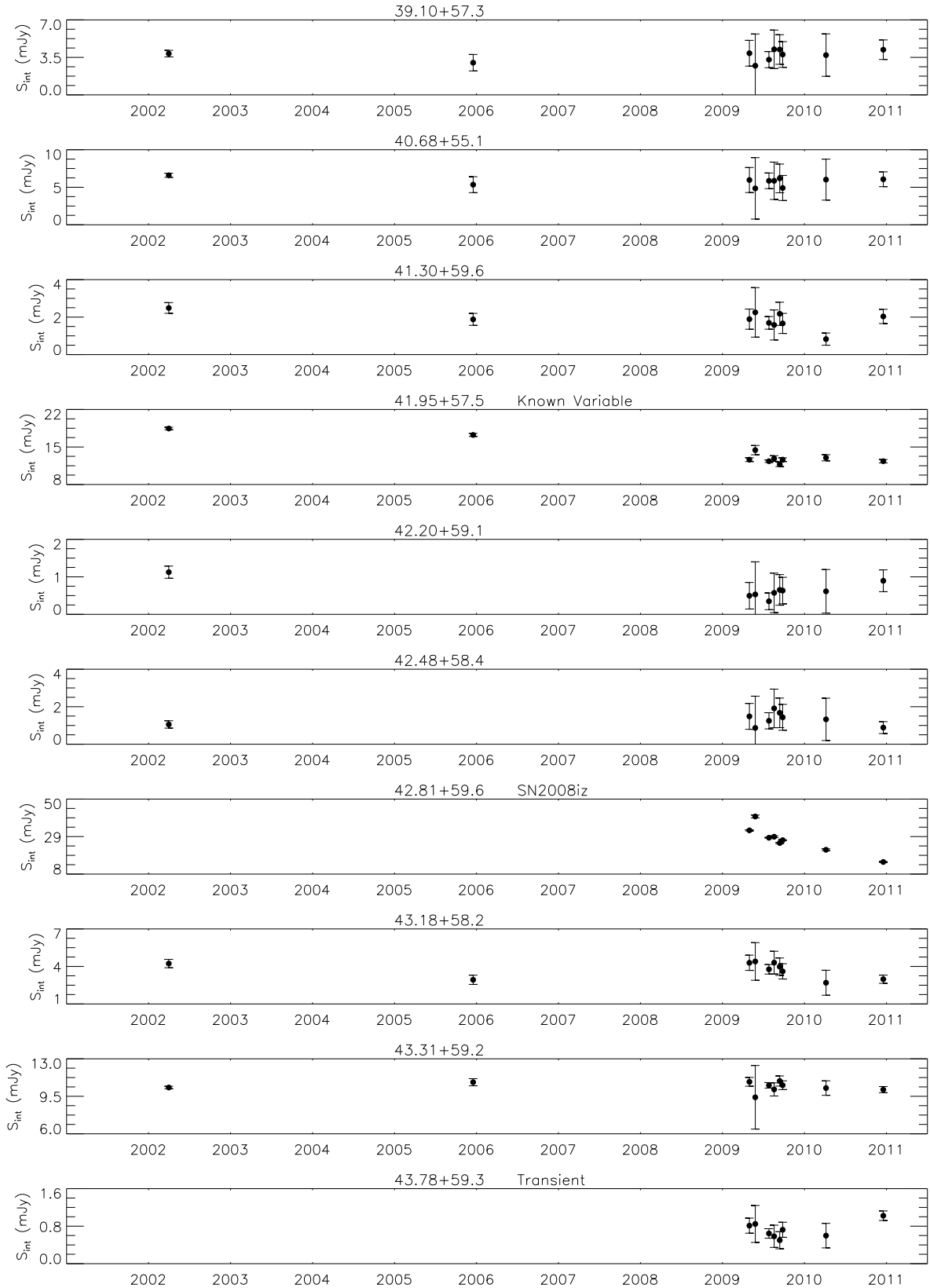}}
    \caption[]{\label{LightCurve}MERLIN and eMERLIN light curves for some sources in M82. The 2002 and 2005 data were retrieved from \cite{Fenech08} and \cite{Argo} respectively. Flux densities have been normalised as described in \S\ref{FMeas}. } 
\end{figure*}
\begin{figure*}
    \centerline{
      \includegraphics[scale=0.9]{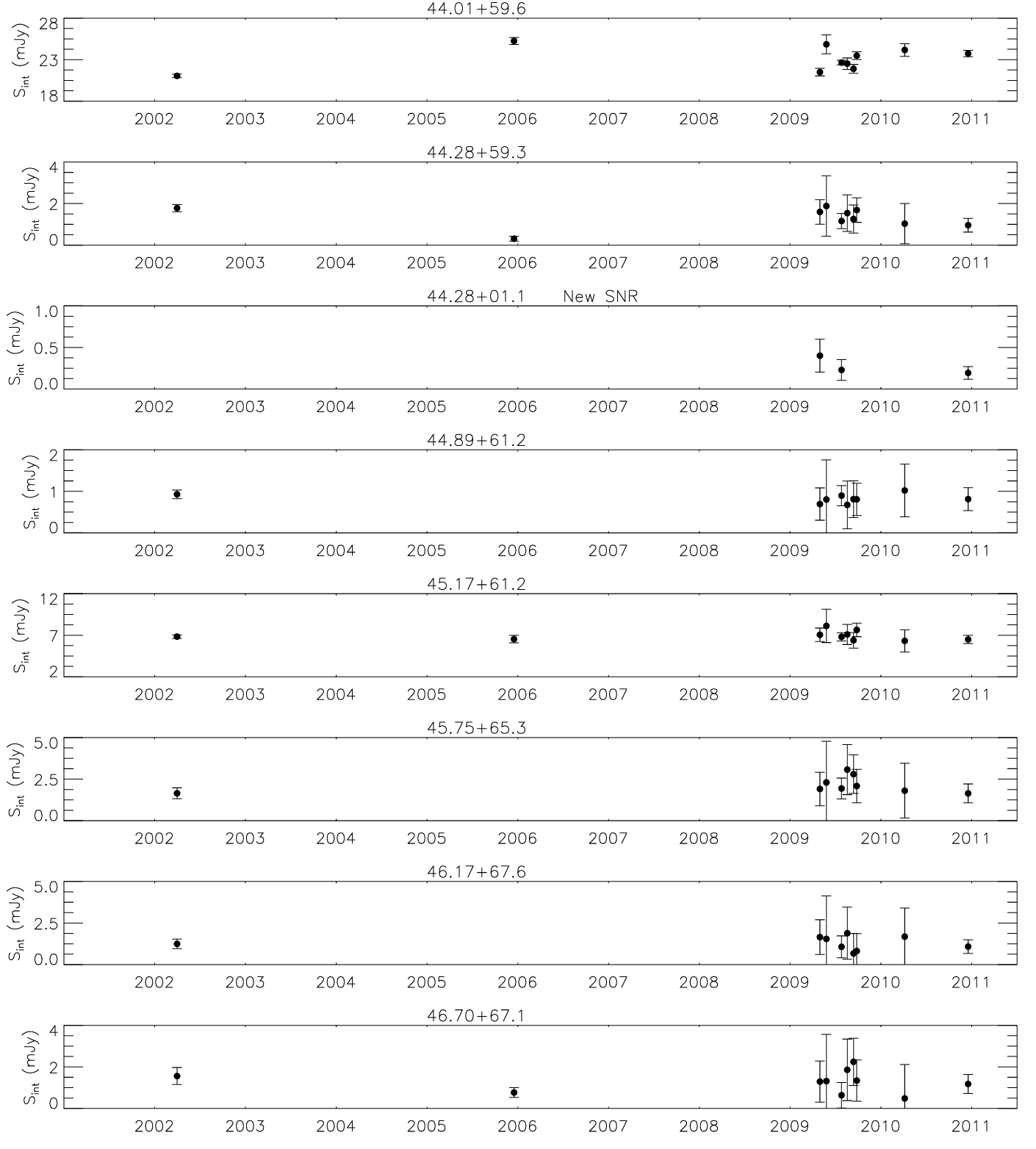}}
     {\bf Figure \ref{LightCurve} - } {\it continued}
\end{figure*}

\subsection{Source identification}

\cite{Fenech08} identified 55 sources during the 2002 deep-integration MERLIN observation of M82, and 49 of them, with $S_{peak} \ge$ 99$\mu$Jy/bm ($\sim 3\sigma$), were detected in our 2009-2010 combined field. Of the other six, three sources lie outside of the imaged area (37.53+53.2, 38.76+53.5 and 46.56+73.8) and three have $S_{peak}< 3\sigma$ (42.62+59.9, 42.66+51.6 and 43.55+60.0). Supernova SN2008iz \citep{Brun09} and the transient source 43.78+59.3 \citep{Mux10} were also identified.

In order to search for possible new sources, each of the 51 identified sources were imaged and subtracted from the uv-data. The resulting dataset was then re-imaged, using the same parameters as the original field. Following inspection of this new image, one new source was identified (44.28+61.1, see \S\ref{PartSource}), bringing the total number of identified sources to 52 (Figure~\ref{M82Panel} and Table~\ref{FluxTab}).

\subsection{Flux density measurements}\label{FMeas}

The flux densities of all compact sources in M82, as well as of the phase calibrator 0955+697, were measured for each of the eight epochs using the \textsc{aips} task \textsc{tvstat}, and any background flux density surrounding the sources subtracted. In order to convert all flux densities in this work to 5\,GHz, spectral index information for each source was retrieved from \cite{McD02}, \cite{Brun09} and \cite{Mux10}. When unavailable, standard values of $\alpha = 0.0$ and $\alpha = -0.7$ were used for \ion{H}{ii} regions and SNR respectively. 

The overall flux density calibration scale, as derived from observations of the standard flux density calibrator source 3C\,286 on an epoch by epoch basis, is accurate to few percent, however some small systematic variations in the flux density scale may remain between individual epochs. Since the phase reference calibrator shows significant flux density variations on the timescales of this monitoring project (Figure~\ref{PHcalflux}), an internal cross-calibration procedure following the methodology of \cite{Kro00} was applied to the results to assess and correct for any systematic differences in the flux density calibration of these data. Following the initial calibration and imaging of these target data, the flux densities of the brightest and most historically stable SNRs identified by \cite{Kro00} were measured to assess whether any systematic, inter-epoch trends were visible in their flux density. 

The bright sources 40.68+55.1, 43.31+59.2, 44.01+59.6 and 45.17+61.2 which showed a consistent trend \citep[as defined by][]{Kro00} were used to provide a re-normalisation of the flux density scale between individual epochs (Figure~\ref{Norm}). For each source, the differences between flux density measurement at a given epoch and the best-fit line to the data is computed. The resulting corrections for each epoch was computed by averaging these differences between each source, epoch by epoch.\\

Details of the averaged  flux density measurements (over the seven 2009 MERLIN epochs) are given in Table~\ref{FluxTab}. In addition to the 2002 \citep{Fenech08} and 2009 data, flux density measurements (which were also normalised as described above) from a 2005 MERLIN monitoring session \citep{Argo} were also included. These observations were conducted at a frequency of 4.754\,GHz in December 2005, with parallel hands of circular polarisation, a bandwidth of  16\,MHz divided into 32 channels, and a total on-source integration time of eight hours.

Data for each epoch can be found in Appendix~\ref{FEpoch} and a few select light curves are displayed in Figure~\ref{LightCurve}.

\begin{figure*}
  \begin{minipage}{16.cm}
    \begin{minipage}{5cm}
      \centerline{
        \includegraphics[scale=0.28]{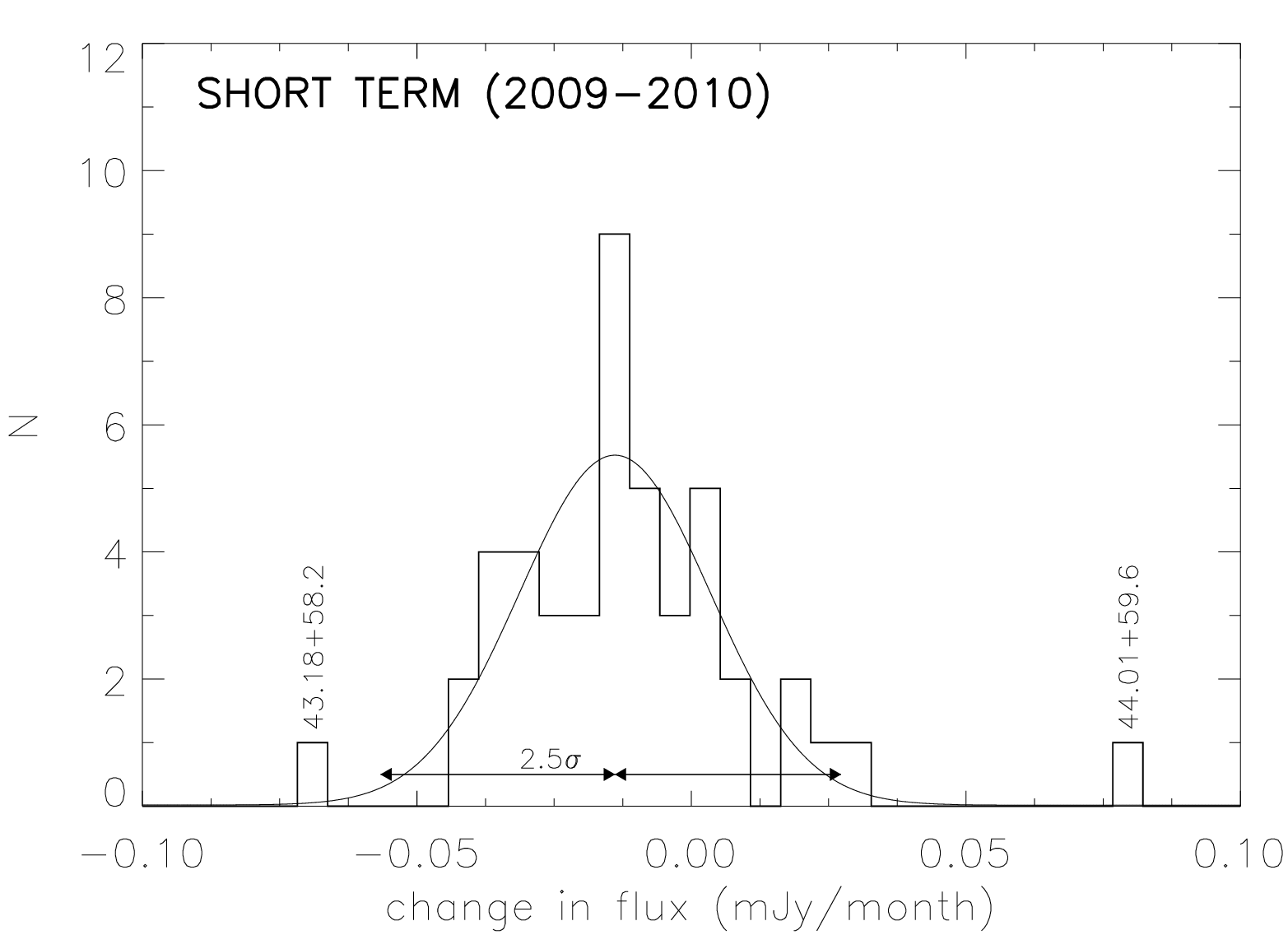}}
    \end{minipage}
    \hfill
    \begin{minipage}{5cm}
      \centerline{
        \includegraphics[scale=0.28]{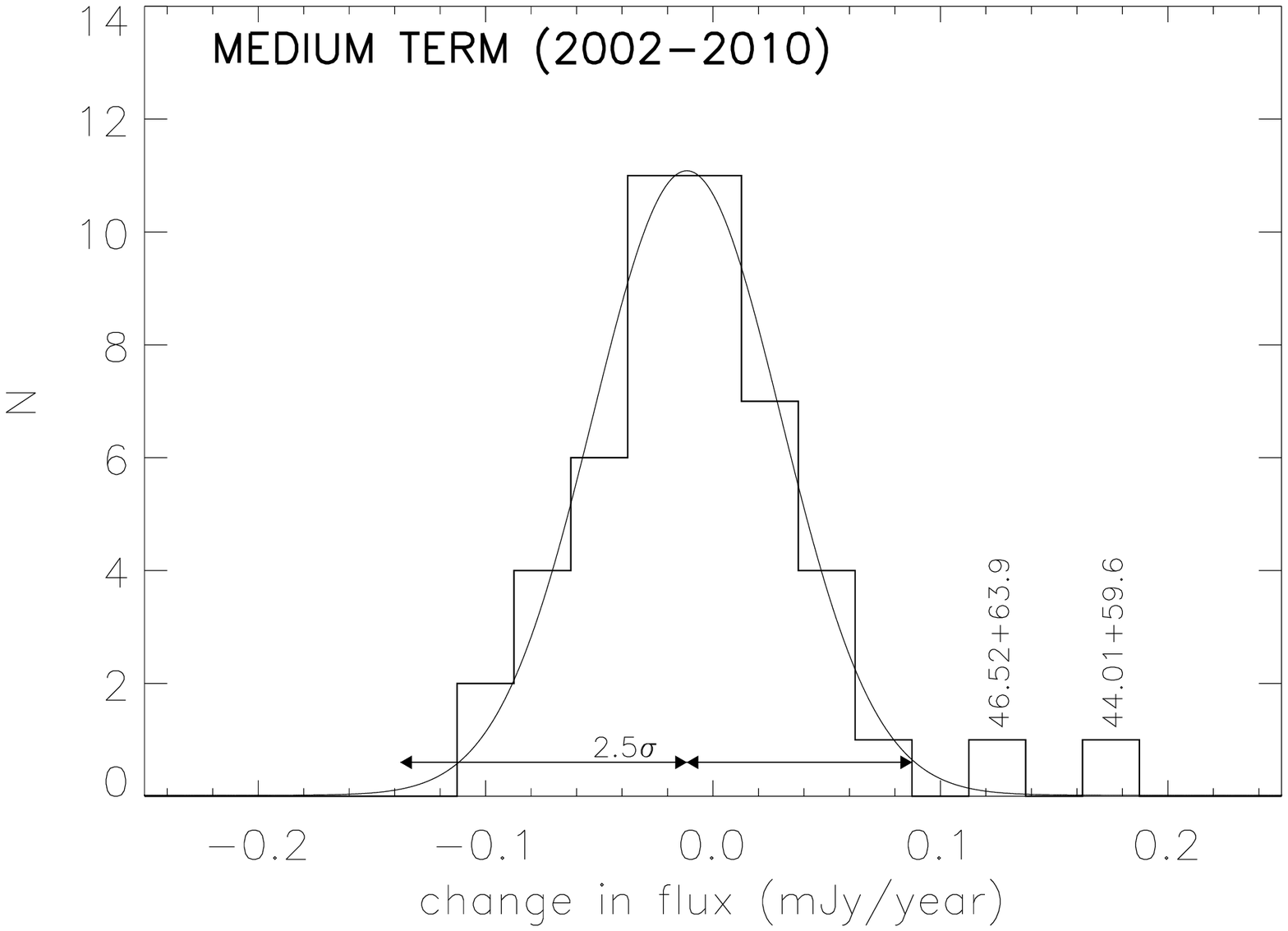}}
    \end{minipage}
    \hfill
    \begin{minipage}{5cm}
      \centerline{
        \includegraphics[scale=0.28]{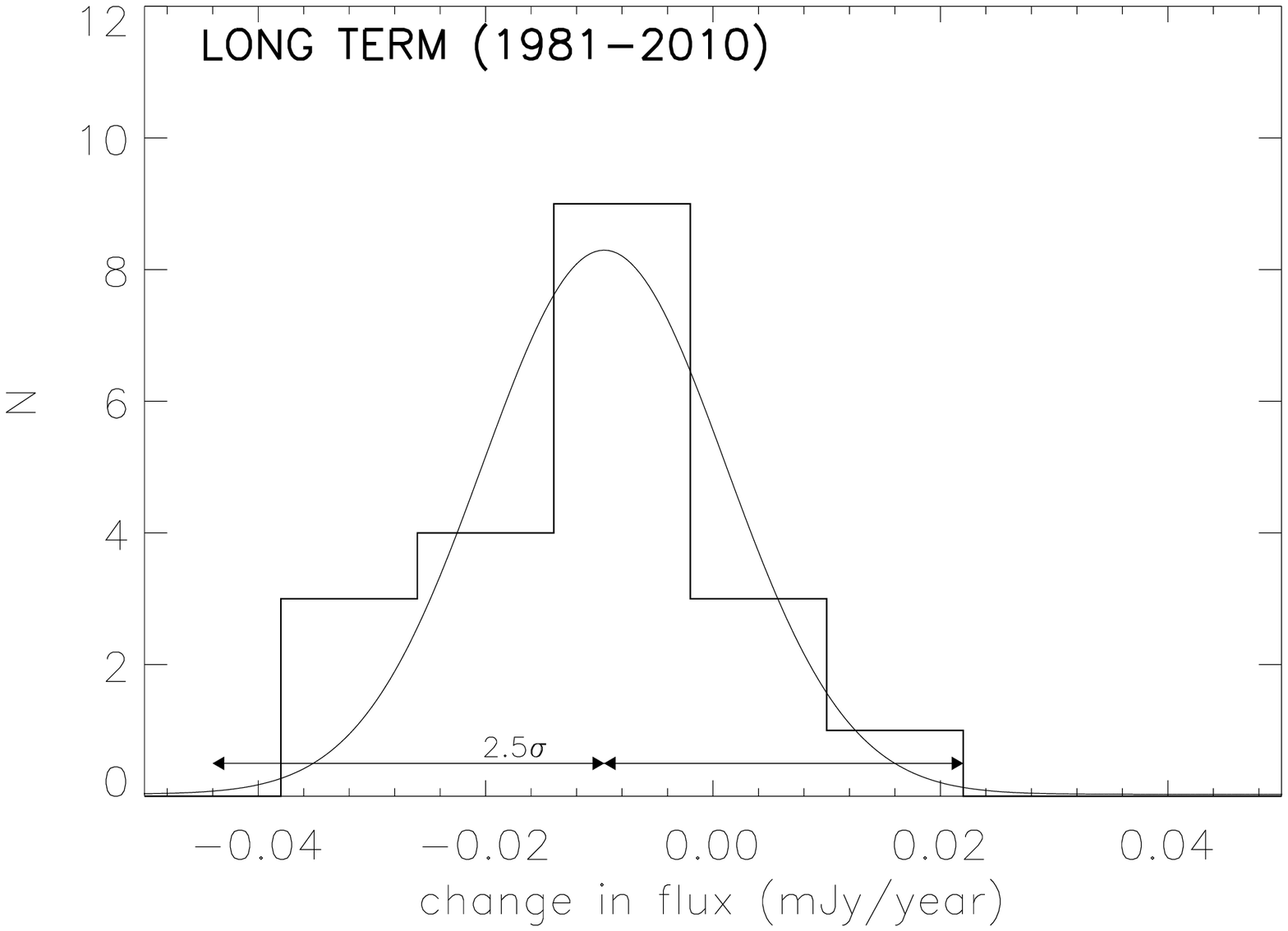}}
    \end{minipage}
  \end{minipage}
    \vspace{2mm}
  \caption[]{\label{RatioS}Short (over the 2009-2010 period, left), medium (over the 2002-2010 period, centre) and long (over the 1981-2010 period, right) term variability, as described in \S\ref{Results}. Names correspond to sources with rates of change outside a 2.5$\sigma$ limit of the distribution. SN2008iz and the variable source 41.95+57.5 lie well outside the range of this plot and are not represented.\\}
\end{figure*}
\begin{figure*}
  \begin{minipage}{10.0cm}
    \begin{minipage}{10.0cm}
      \centerline{
        \includegraphics[scale=0.5]{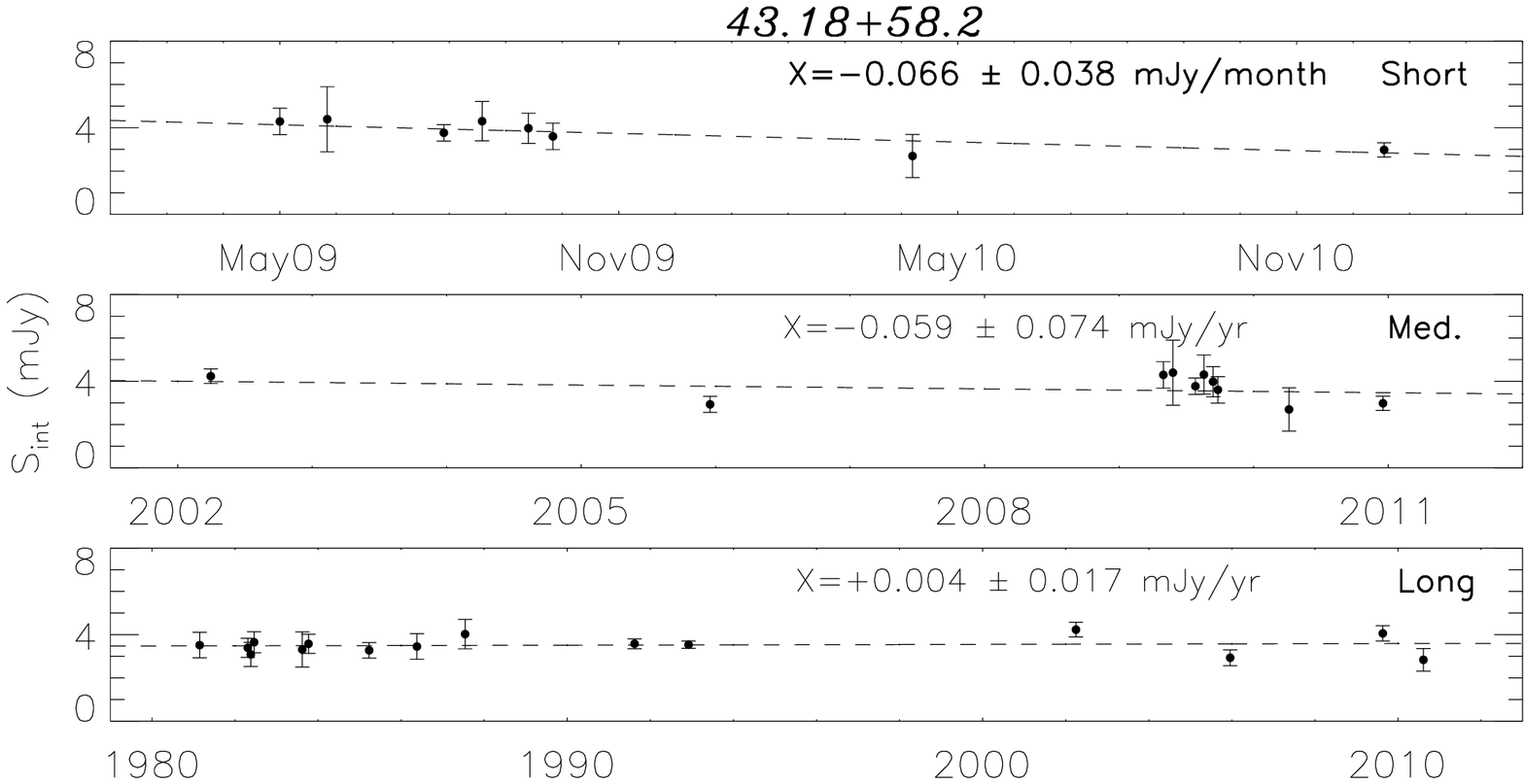}}
    \end{minipage}
    \vfill
    \vspace{5mm}
    \begin{minipage}{10.0cm}
      \centerline{
        \includegraphics[scale=0.5]{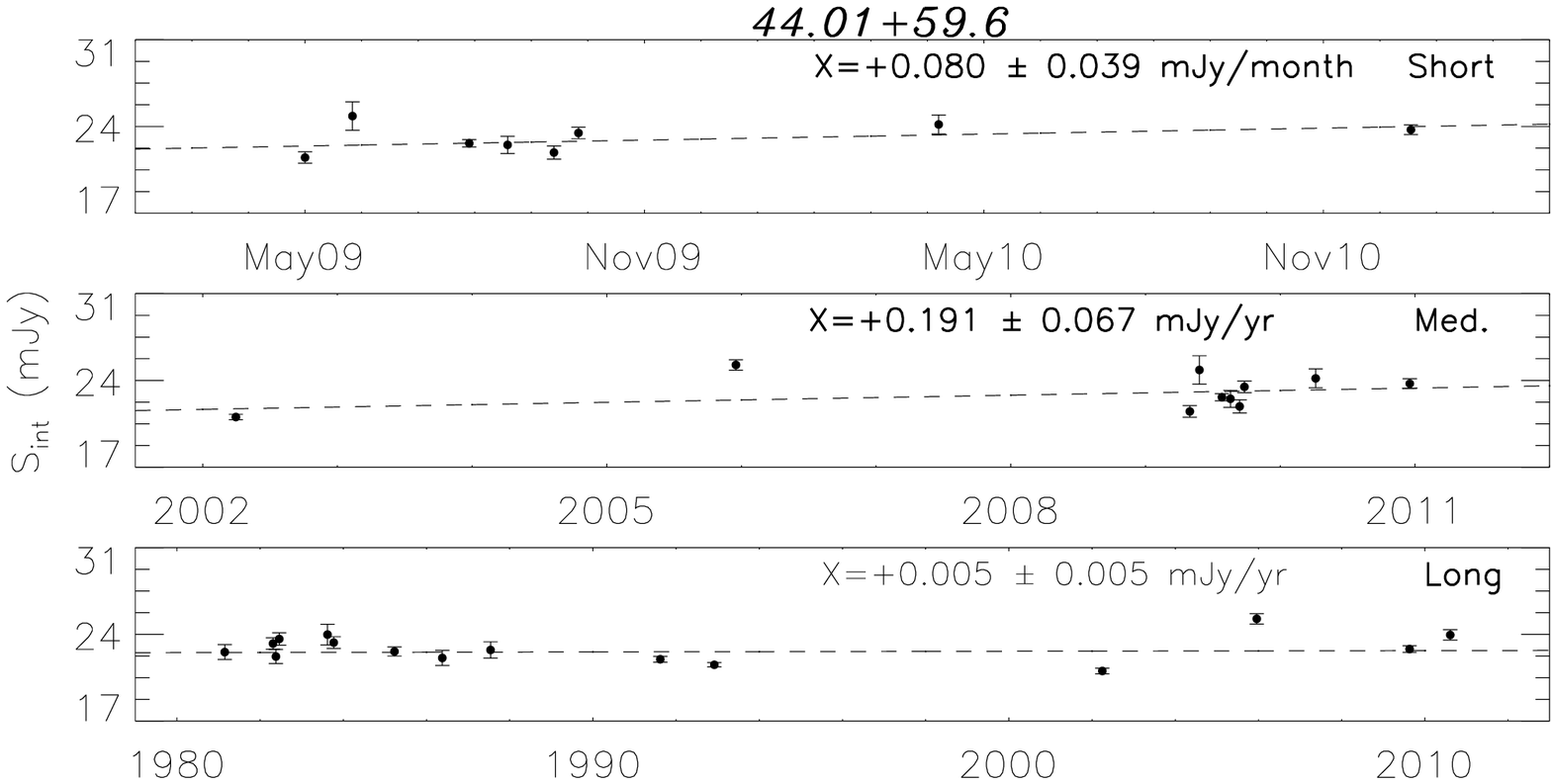}}
    \end{minipage}
    \vfill
    \vspace{5mm}
    \begin{minipage}{10.0cm}
      \centerline{
        \includegraphics[scale=0.5]{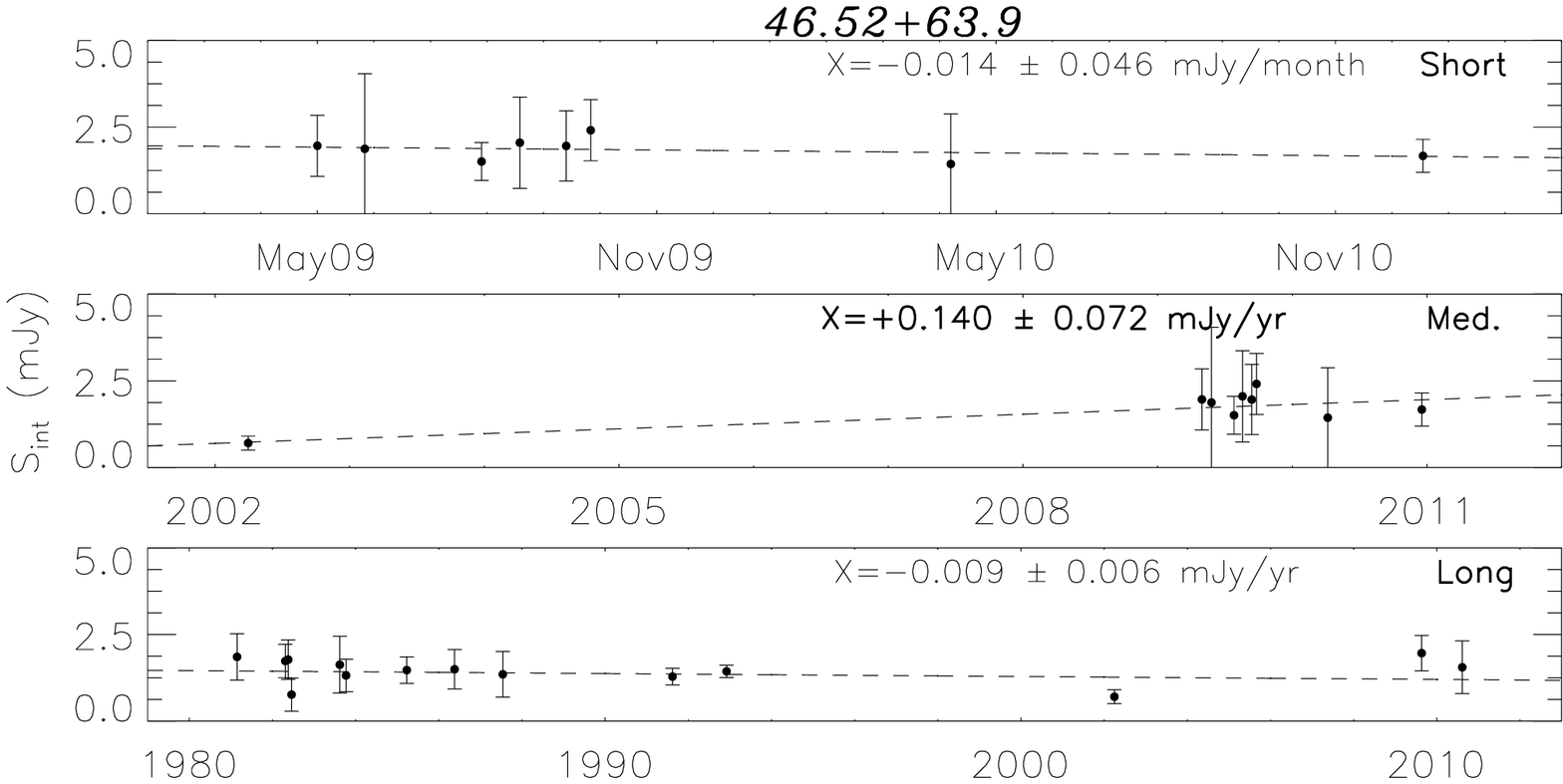}}
    \end{minipage}
  \end{minipage}
  \begin{minipage}{5.0cm}
    \begin{minipage}{5.0cm}
      \centerline{
        \includegraphics[angle=270,scale=0.25]{Figures/43.18+58.2_2009.ps}}
    \end{minipage}
    \vfill
    \vspace{7mm}
    \begin{minipage}{5.0cm}
      \centerline{
        \includegraphics[angle=270,scale=0.25]{Figures/44.01+59.6_2009.ps}}
    \end{minipage}
    \vfill
    \vspace{7mm}
    \begin{minipage}{5.0cm}
      \centerline{
        \includegraphics[angle=270,scale=0.25]{Figures/46.52+63.9_2009.ps}}
    \end{minipage}
  \end{minipage}
  \hspace{5mm}
  \caption[]{\label{DecSour}Left: Short, medium and long term light curves over the period 2009-2010 (top), 2002-2010 (centre) and 1981-2010 \citep[bottom, where the 1981-1991 data were retrieved from][]{Kro00} for sources showing noticeable decrease in flux density (shown in bold font on the plots), aside from 41.95+57.5 and SN2008iz (Table~\ref{DecSourTab}). For long term variation, flux measurements were averaged over EP1-6 for 2009 and EP7-8 for 2010 (as described in Table~\ref{EpochTab}). A line $S(t)=S_0+Xt$ was fitted to each curve (dashed line), where X characterises the variation in mJy/month and mJy/yr for short and medium/long terms respectively. Right: Contour plot at $-$1, 1, 1.414, 2, 2.828, 4, 5.656, 8, 11.282, 16, 22.564, 32, 45.1286 $\times$ 99 $\mu$Jy/bm of the source from the combined 2009-2010 MERLIN data. The beam size of 50$\times$50 mas is shown in the bottom left-hand panel.\\} 
\end{figure*}
\begin{table*}
  \begin{center}
    \caption[]{\label{DecSourTab}Characteristics of the sources showing flux density variations (shown in large font in the table), aside from 41.95+57.5 and SN2008iz, in M82. Short term decrease rates were derived from the 2009-2010 data while sizes, radial velocities and ages were retrieved from \cite{Fenech08}$^{[1]}$. Medium term decrease rates were computed over the period 2002-2010 and long term decrease rates over 1981-2010, where the 1981-1991 data were retrieved from \cite{Kro00}. The corresponding light curves are shown in Figure~\ref{DecSour}.}
    \medskip
    \begin{tabular}{cccccccc}
      \hline
             & 2009 Integ. &\multicolumn{3}{c}{Apparent rate of change} & Size$^{[1]}$   & Radial  & Age$^{[1]}$    \\
      Source & Flux Dens.&{\bf Short Term} &  {\bf Medium Term}  & {\bf Long Term}    &  & Velocity$^{[1]}$&  \\
      &&{\bf 2009-2010}& {\bf 2002-2010}& {\bf 1981-2010}  &  & & \\
             & (mJy) &(mJy/month)  & (mJy/yr)  & (mJy/yr)      & (pc)    & (km/s)  & (yr)\\
      \hline
43.18+58.2 & \phantom{1}4.79$\pm$0.33 & {\large $-$0.066$\pm$0.038} &         $-$0.059$\pm$0.074  &   +0.004$\pm$0.017 & 1.9$\pm$0.1 & 10500$\pm$2300            & \phantom{1}55\\
44.01+59.6 &           25.88$\pm$0.25 &  {\large  +0.080$\pm$0.039} &   {\large +0.191$\pm$0.067} &   +0.005$\pm$0.005 & 0.8$\pm$0.1 & \phantom{1}2700$\pm$400\phantom{0}& 140\\
46.52+63.9 & \phantom{1}2.68$\pm$0.47 &         $-$0.014$\pm$0.046  &   {\large +0.140$\pm$0.072} & $-$0.009$\pm$0.006 &      -      &           -              & - \\
      \hline
    \end{tabular}
  \end{center}
\end{table*}

\section{Results}\label{Results}

Since very rapid changes in flux density are not physically expected for the observed sources, and epoch-to-epoch variations can be strongly dependent on calibration methods, the approach to measure variability used in this work does not examine these very short-term changes. With these considerations in mind, we find that most of the sampled sources in M82 show no significant changes in flux densities over all timescales.\\

To improve our analysis of variation, data taken over an 11-year period from 1981 \citep{Kro00} were added to the light curves when available (21 out of 52 sources). A systematic shift of $\sim$1.07 mJy between the VLA of \cite{Kro00} and MERLIN data was removed from the VLA data. This difference of the overall flux density of each source is due to the higher beam size of the VLA observation of \cite{Kro00}, which will be sensitive to more diffuse matter than the MERLIN data presented here. Changes in the flux density of the detected sources were studied in the short, medium and long term (over the 2009-2010, 2002-2010 and 1981-2010 periods respectively). A line S(t) (with t in months for short term and years for medium and long terms) was fitted through every light curve and the slope used to determine which sources displayed consistent flux density variations. The distribution of slopes was fitted with a Gaussian model (Figure~\ref{RatioS}). Outliers (with rate of change outside of 2.5$\sigma$) indicate sources with noticeable variability. Note that SN2008iz (X$_{short}$=$-$0.9 mJy/month) and the variable source 41.95+57.5 (X$_{long}$=$-$3.1 mJy/year), which show extreme variation, are not represented in these plots. In addition, the Gaussian distributions in Figure~\ref{RatioS} show an offset from zero, possibly showing a consistent decrease in flux densities of all sources at all epochs, although within one sigma of the measurements.

The left panel of Figure~\ref{RatioS} shows flux variability in the short term (over the period 2009-2010). The sources 43.18+58.2 and 44.01+59.6 show decreases in flux density. 
The central panel of Figure~\ref{RatioS} shows flux density variability in the medium term (over the period 2002-2010). The sources 44.01+59. and 46.52+63.9 show steady changes in flux density. 
Finally, the right panel of Figure~\ref{RatioS} shows flux variability in the long term (over the period 1981-2010). No evidence of systematic variation in flux density is seen in any of the 21 sources considered. 
All variability values are shown in Table~\ref{DecSourTab}, while the light curves for varying sources are displayed in Figure~\ref{DecSour}.

We note that most of the sources showing short term flux density variation are the most compact objects on VLBI scales in M82 \citep{Fenech08} after 41.95+57.5, whose flux-density is known to steadily decrease by $\sim$8.8\%/yr \citep{Trot96,Kro00,Mux05}. Each source showing possible flux density variations is discussed in more detail below.

\subsection{Variable sources}\label{PartSource}

A total of six sources have shown flux density changes (i.e. with rate of change outside of 2.5$\sigma$ of the distribution) in the short, medium or long term, including 41.95+57.5, SN2008iz and the transient source 43.78+59.3. The latter, being quite particular objects, will be discussed separately in Sections \ref{41.95}, \ref{SN2008iz} and \ref{Trans}. Some characteristics of the other sources, including flux density rates of change and sizes, are given in Table~\ref{DecSourTab}.

As previously stated, two of the three other varying sources (43.18+58.2 and 44.01+59.6) have small angular sizes, and are in fact some of the most compact SNR in M82 after 41.95+57.5. Based on expansion velocity measurements from VLBI and MERLIN data, \cite{Fenech08} and \cite{Besw06} determined that these sources are amongst the youngest in M82, with ages of 55 and 140 yrs (as of 2002). Consequently, their short and medium term brightness variations could be explained by changes in the circumstellar and interstellar medium through which these shocks travel. This may imply that these sources are at a different temporal stage in their evolution compared with the older more stable sources.

Finally, the third source showing flux density variation, 46.52+63.9, shows only medium term variability. From its light curves in Figure~\ref{DecSour}, the variation can be explained by an anomalously low value of the source flux density measured in 2002. The source is then excluded from the varying sample.

\subsubsection{41.95+57.5}\label{41.95}

This source has shown a continued decrease in flux density of $\sim$8.8\%/yr since its first observation in 1965 \citep{Trot96,Mux05}. This decay rate would imply that, at birth, the source would have had a flux density of $\sim$30 Jy \citep[assuming a free expansion and a source age of $\sim$80 years;][]{Fenech08}. These two facts (continuous decay and high flux density at birth) suggest that 41.95+57.5 is likely to be an exotic event and several suggestions have been made as to its nature, including the possibility that it may have been an off-axis gamma-ray burst \citep{Mux05}.

\begin{figure}
  \centerline{
    \includegraphics[scale=0.4]{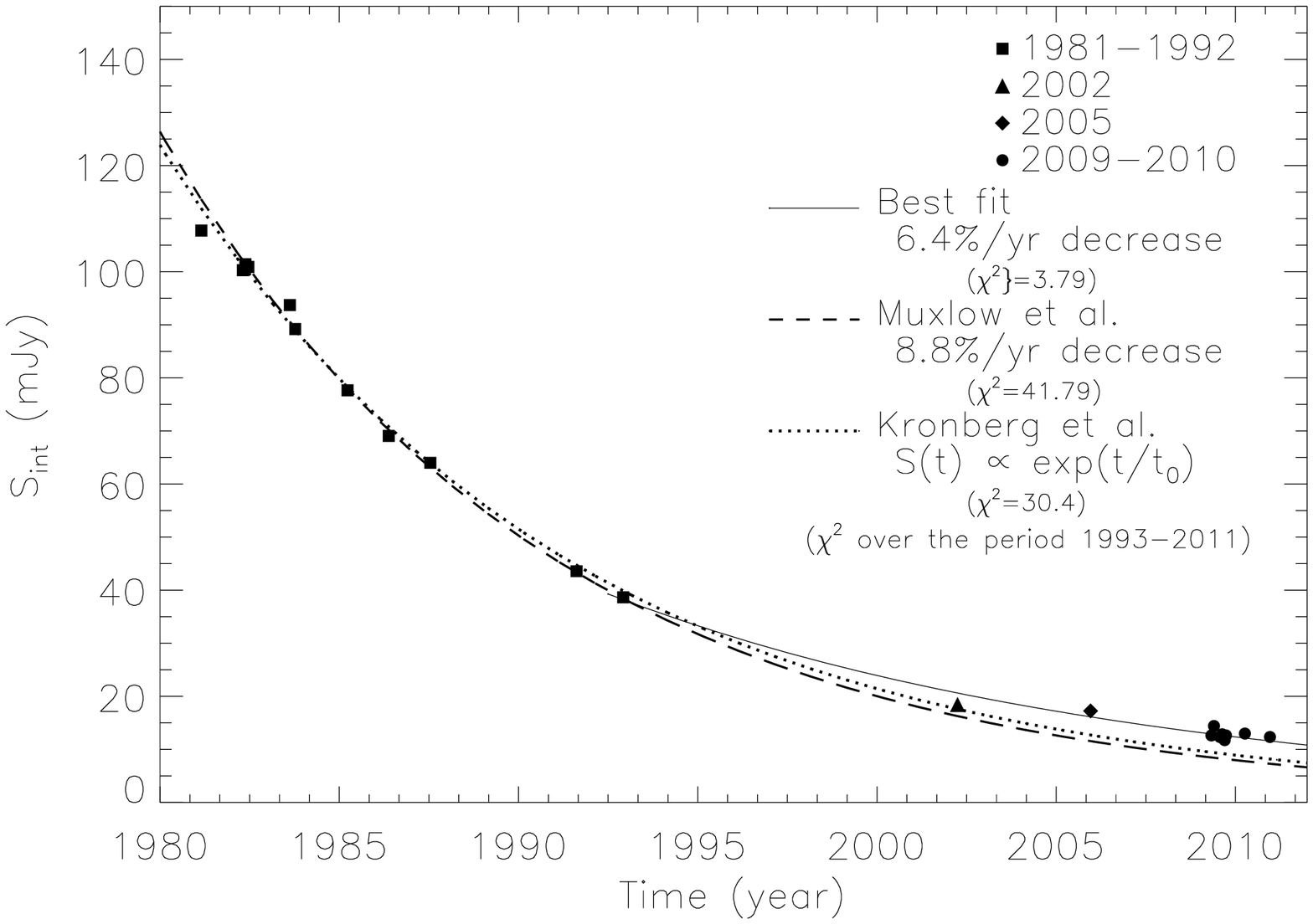}}
  \caption[]{\label{Kro4195}Light curve for the source 41.95+57.5 over the period 1981-2010. Data for each period were retrieved as follow: 1981-1991 from \cite{Kro00}, 2002 from \cite{Fenech08}, 2005 from \cite{Argo}, 2009-2010 from this work. Models for the flux density decrease are based on an exponential decay \citep[dotted line;][]{Kro00}, an 8.8\%/yr decay \citep[dashed line;][]{Mux05} and our best fit 6.4\%/yr decay (solid line) for the period 1993-2010. The values of $\chi^2$ quoted here correspond to the reduced chi-square statistics for each fit.}
\end{figure}

As seen in Figure~\ref{Kro4195}, 41.95+57.5 still shows a steady decrease in flux density. However, its rate of change in flux density has decreased from 8.8\%/yr to 6.4\%/yr in 1981 to 6.4\% in 2010, when fitting between 1993-2010.

VLBI observations of 41.95+57.5 by \cite{Besw06} in November 1998 and February 2001 shows the source to be expanding at a rate of $\sim$2000$\pm$500 km s$^{-1}$. However, further VLBI data from 2005 \citep{Fenech07,Fenech10} shows that the apparent radius of 41.95+57.5 may have decreased between 2001 and 2005 (30.4 mas in 2001 and 26.85 mas in 2005).\\

The unusually luminous radio supernova SN1986J in NGC891 and 41.95+57.5 show some distinct similarities. Both sources display an asymmetric radio structure on milliarcsecond scales, modest (few 1000\,km\,s$^{-1}$) expansion velocities, relatively high initial radio luminosities and a power-law flux density decay \citep{Biet10}. In recent VLA and VLBI observations the spectral index of SN1986J has been seen to be evolving rapidly above 5\,GHz, deviating from the flux density decline observed at lower frequencies. This has coincided with 8.4\,GHz and 15\,GHz global VLBI observations detecting an emerging, inverted spectrum source in the centre of the SNR \citep{Bart09}. It has been asserted that this new source may be a plerion, or pulsar wind-nebula \citep{Biet04}.

Both the possible decrease in apparent angular size \citep{Besw06,Fenech08} and the small deviation from the historical flux density decline of 41.95+57.5 in these recent 5\,GHz flux density measurements could be consistent with the potential appearance of a new higher frequency central component within the shell, or that 41.95+57.5 comprises of two components: a central, constant flux density one with a thermal-like spectrum, plus a larger component with a synchrotron spectrum that is simultaneously expanding and declining in flux density. In both of these scenarios, up until recent times, the synchrotron source has dominated the observed characteristics of the object. As such, this source may be undergoing an evolution analogous to that observed in SN1986J, albeit in a much older remnant. Unfortunately, the MERLIN and e-MERLIN observations presented here do not have adequate angular resolution to image in detail the structure of this source and, as yet, higher frequency ($>$1.6\,GHz) VLBI observations have not confirmed this hypothesis. Further analysis of the variation in flux densities of 41.95+57.5 and SN1986J, both in time and frequency, would be needed to assess the similarity between the sources.

\subsubsection{42.81+59.6 - SN2008iz}\label{SN2008iz}

\begin{figure}
  \centerline{
    \includegraphics[scale=0.42]{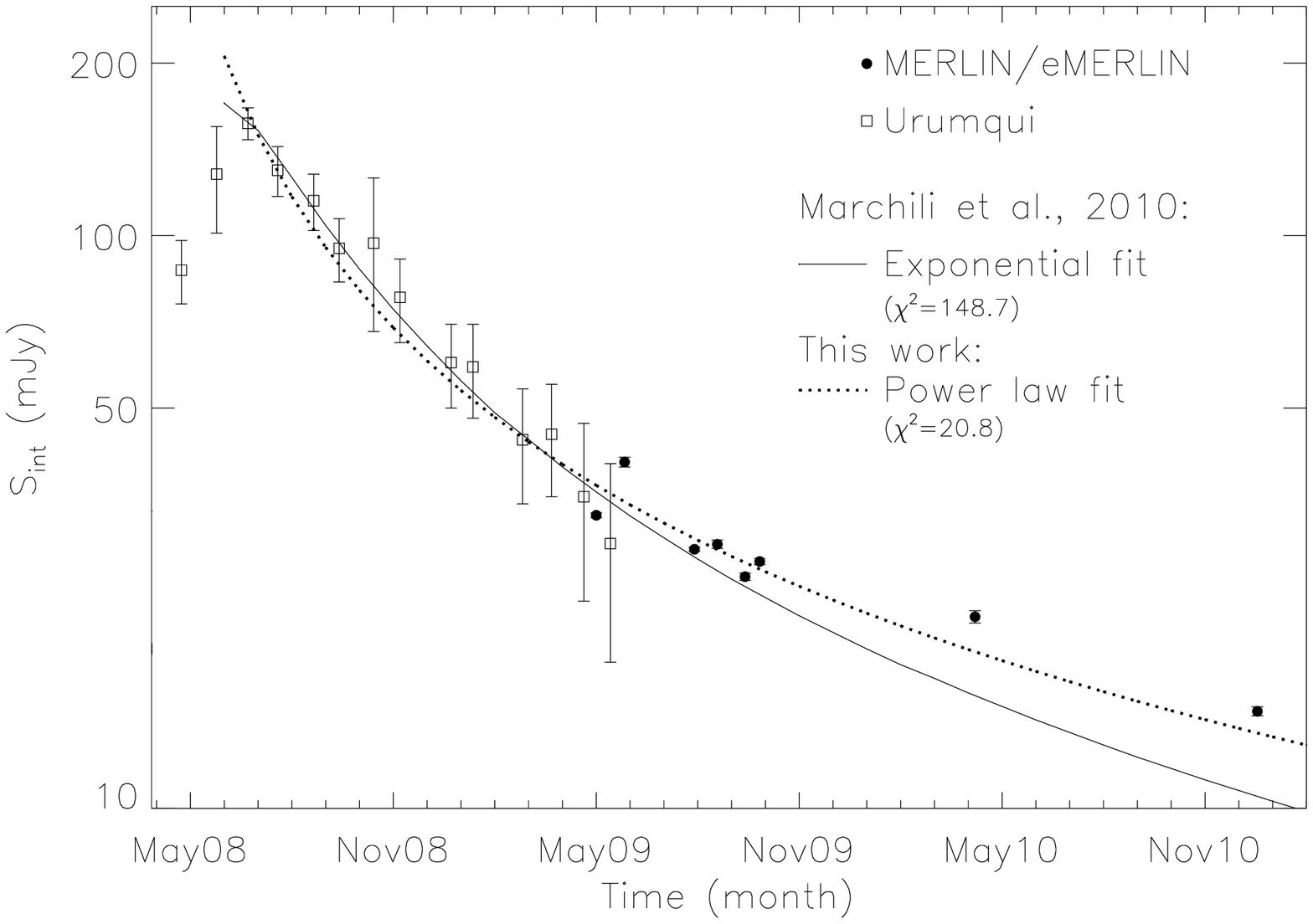}}
  \caption[]{\label{SNPic}Light curve for SN2008iz. 5\,GHz data from single dish Urumqi observations \citep{March10} and the MERLIN/{\it e}-MERLIN data are shown as open squares and filled circles respectively. An exponential decay model (Equ.~\ref{explaw}) was fitted to the curve using parameters from \cite{March10} - solid curve - as well as a power-law decay model - dotted curve - which displays the best fit to the decay part of the curve. The values of $\chi^2$ quoted here correspond to the reduced chi-square statistics for each fit.} 
\end{figure}

The light curve of SN2008iz from the presented MERLIN data and supplemented by data from \cite{March10}, is shown in Figure~\ref{SNPic}. Following \cite{March10}, the flux density decline after reaching peak brightness can be modelled using an exponential decay,
\begin{eqnarray}\label{explaw}
  S(t)=K_1(t-t_0)^{\beta}e^{-\tau}   \ \ \ \ \ \ \ \ \  \tau = K_2(t-t_0)^{\delta}
\end{eqnarray}
where $t_0$ is the explosion date, which was fitted to be February 18, 2008 \citep{March10}. 

The 2009-2010 data display a flattening of the light-curve which does not seem to be properly fitted by an exponential model. A power-law decay of the form 
\begin{eqnarray}\label{powlaw}
  S(t) \propto (t-t_0)^{\alpha}
\end{eqnarray}
where $\alpha$ = $-$1.15$\pm$0.05, shows a better fit to the flux-density decrease, with a $\chi^2_{red}$=20.8 while the exponential fit of \cite{March10}, including the 2009-2010 data, yields $\chi^2_{red}$=148.7.

Whilst further later time measurements are required to confirm these results, there is tentative evidence that the radio light curve of SN2008iz is showing a small reduction in the rate of its flux density decline.

\subsubsection{43.78+59.3 - Transient source}\label{Trans}

This source was not present in the 2002 and 2005 data but was detected in 2009 \citep{Mux10} and in the MERLIN observation presented here. \cite{Mux10} suggested that this object might either be due to accretion around a massive collapsed object or be a faint and unusual supernova \citep{Jos10}. In the latter case, we would expect the flux density to steadily decrease over time, as the source would be too young to have reached the remnant phase. However, in the observations presented here, it shows a flux density at 5\,GHz of $\sim$0.90 mJy, which does not seem to vary significantly over the period May 2009 - December 2010. The supernova hypothesis is thus unlikely.

\subsection{44.28+61.1- A newly detected SNR shell}\label{NewSource}

\begin{figure}
  \begin{minipage}{8.0cm}
    \centerline{
      \includegraphics[angle=270,scale=0.3]{Figures/44.28+61.1_shell.ps}}
    \caption[]{\label{ShellPic}Contour plot at $-$1, 1, 1.414, 2, 2.828, 4, 5.656, 8, 11.282 $\times$ 99 $\mu$Jy/bm of SNR 44.28+61.1. The beam size of 50$\times$50 mas is shown in the bottom left-hand panel.} 
  \end{minipage}
  \vfill
  \vspace{2mm}
  \begin{minipage}{8.0cm}
    \centerline{
      \includegraphics[scale=0.4]{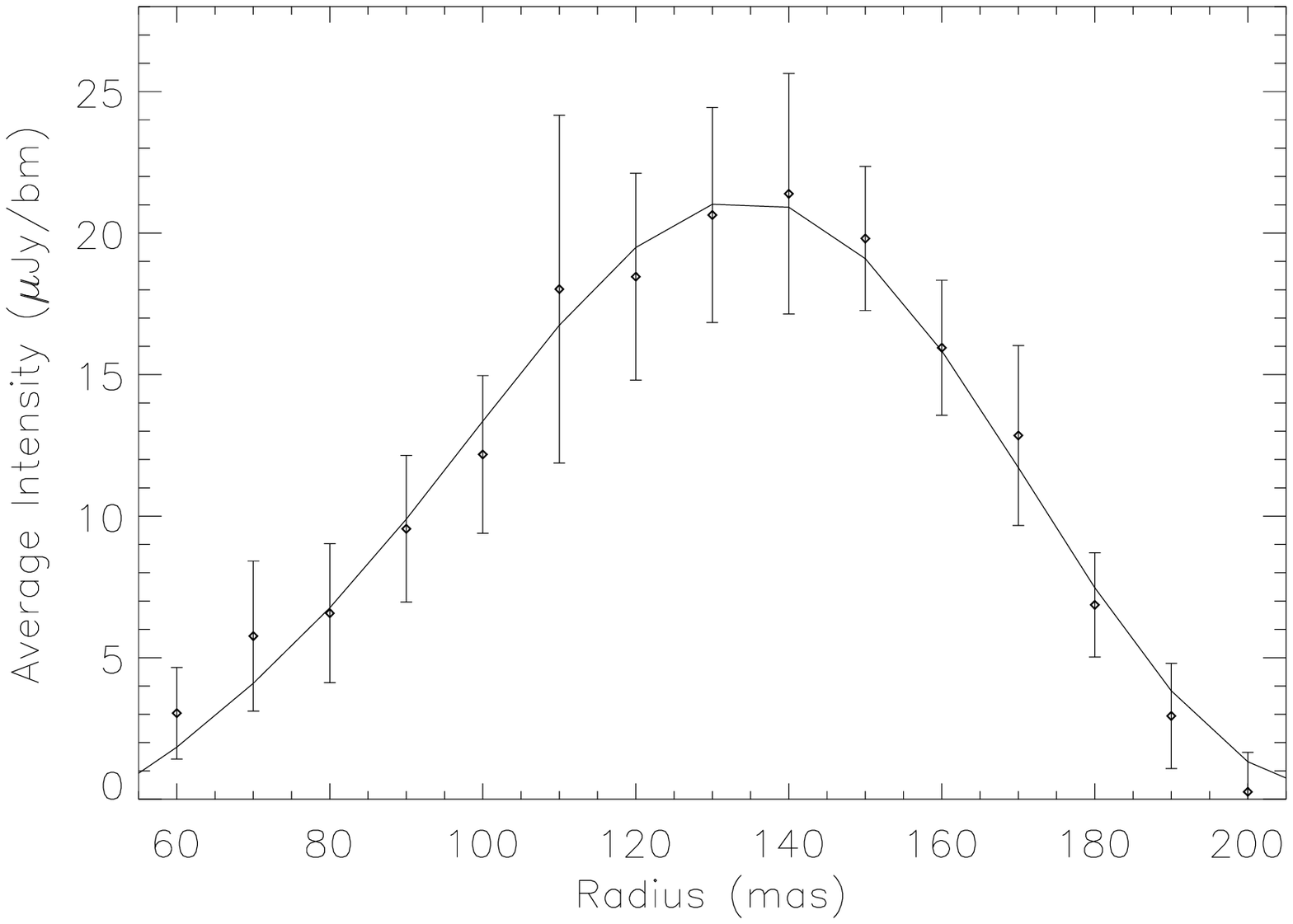}}
    \caption[]{\label{ShellIring}Azimuthally averaged radial intensity profile and best-fit model (solid line) of SNR 44.28+61.1 obtained from \textsc{aips} annular profiles task \textsc{iring}. The peak intensity occurs at {\it r}=140$\pm$5 mas, taken to be the radius of the shell.} 
  \end{minipage}
\end{figure}

This source, located at position 09h55m53.00s +69$^{\circ}$40$^{\prime}$47.24$^{\prime\prime}$ (J2000), was not visible in the 2002 MERLIN data. It has a peak flux density value of $S_{peak}$=0.18 mJy/bm and an integrated flux density of 0.40$\pm$0.36 mJy from the combined field. Being shell like in shape (Figure~\ref{ShellPic}), it is most likely a supernova remnant. 

By measuring the source extent, the age of the SNR can be inferred. The radius of the source was determined using the \textsc{aips} annular profiles task \textsc{iring}, giving {\it r}=140$\pm$5 mas (Figure~\ref{ShellIring}). At the distance of M82 (3.2 Mpc), this gives an estimated radius of 2.1\,pc. \cite{Fenech08} measured expansion velocities of SNRs in M82 to be between 2200 and 10500 km/s, with a mean velocity of $v_{exp}$ = 5650 km/s. Using the latter value, we can estimate the age of 44.28+61.1 to be $\sim$360\,yrs.


\section{Discussion}

\subsection{Possible observational effects}

Before discussing possible physical explanations for the long-term observed flux density decay of the sources in M82 (or lack thereof), it would be prudent to consider any observational explanations for any apparent variation. The data from \cite{Kro00} are from VLA A-configuration observations of M82 with a resolution of 350\,mas compared to the 50\,mas resolution of the 2009-2010 MERLIN observations. As such, virtually all of the sources discussed that were observed in both sets of observations will be unresolved in these VLA data and resolved in these MERLIN data. As the MERLIN array lacks shorter baselines, it is comparatively insensitive to the larger spatial scales, potentially `resolving out' some of the emission associated with a given SNR. It is however considered unlikely that this would explain the observed flux density decay for the varying sources, and such an effect would become more pronounced for the larger, fainter and more diffuse SNR. All of the larger, diffuse sources do not show any flux density variation. This is compounded by the fact that the decay is not only apparent between the VLA and MERLIN data, but continues within the MERLIN-only data.\\

The final observational consideration is that we are essentially measuring structurally evolving sources with a fixed system i.e. with a fixed uv-distribution which therefore becomes sensitive to slightly different structures as a source evolves. However it is unlikely that this would account for the flux density decreases observed, particularly over the timescales considered. All of the varying sources are well-defined, in general compact, and bright, thus such an effect would be more observable for higher velocity sources or those that are already more diffuse.

\subsection{Physical explanation}

Of the sources observed in the data presented here, two show short and medium term variations in flux density: 43.18+58.2 and 44.01+59.6. These sources are some of the most compact objects in M82 after 41.95+57.5, SN2008iz and the transient source 43.78+59.3, with measured sizes of 0.8 and 1.9\,pc in MERLIN 2002 observations \citep{Fenech08}.\\

As discussed in \cite{Fenech10}, the SNRs in M82 are expected to be the result of core-collapse supernova events with a massive, typically red supergiant (RSG), progenitor star. At the time of the supernova explosion, there will be a complex circumstellar environment consisting of a slow-moving dense RSG wind with a density profile of $\rho \propto r^{-2}$ followed by a wind-blown bubble of almost constant low density produced by the main-sequence low-density, high-velocity wind. Beyond this wind-blown bubble, there will exist potentially the remains of a collapsed H{\small II} region and eventually the ambient ISM \citep[][and references therein]{Wei02}.

It is generally accepted that following the evolution through the wind-blown bubble, it is likely that there will be a peak in flux density as the supernova ejecta travels from the low-density wind-blown bubble into the higher density ISM (particularly in the case of M82). This will primarily be caused by the production of a thin-dense shell at the edge of the wind-blown bubble. There will however, also be a subsequent decline in flux density as the ejecta traverses this shell and emerges into the ISM, traditionally considered the transitional phase from a supernova to a supernova remnant. It is possible that these transition phases could provide an explanation for the observed flux density changes.\\

M82 SNRs appear to be a more luminous population than those in the Milky Way and the LMC, and typically appear to settle into a constant flux evolution on shorter time scales and at smaller diameters than their Galactic counterparts. This is commonly attributed to the higher density environment within the starburst region in which they are situated. The high pressures associated with starburst activity reduces the ability of mass loss from the progenitor star to expand into the ISM. This implies that the denser the environment, the shorter the distance over which a shock has to travel to have fully traversed the circumstellar medium and sweep up sufficient material to enter Sedov evolution and become a well established SNR. More extreme starburst galaxies, such as Arp220, have shown a mixed population of compact sources covering SNe, SNRs and transition objects. They all typically show smaller sizes and higher luminosities, consistent with an extrapolation of evolution within an even denser environment than that found in M82.\\
\indent
A study of SNRs in Arp220 by \cite{Bat11} determined that 3-5 of 18 sources, with sizes between 0.2 and 0.4\,pc, show variations over the period 2006-2008 (i.e. there is a change in flux $>$50\% between the two epochs), corresponding to short timescale changes in this work. These numbers are in accordance with our findings here, considering that Arp220 is a more active starburst galaxy than M82. From figure~5 of \cite{Bat11}, the variable sources in Arp220 and the compact variable sources in M82 fall into the range of sizes which is compatible with an SN to SNR transition (0.2-0.3\,pc in Arp220 and 0.7-2.0\,pc in M82). This progression could explain the changes in flux density.\\
\indent
\cite{Sea07} explore an alternative explanation for the compact non-thermal sources observed in M82, specifically that they could be wind-driven bubbles (WDBs) associated with very young super star clusters as opposed to SNR. One of the primary motivations for presenting this hypothesis is the lack of flux density variability for the majority of the compact sources within M82 as reported by \cite{Kro00}. 
The work presented here has shown that two of the most compact sources (apart from SN2008iz and 41.95+57.5) show flux density variability in the short and/or medium term over the period 2002-2010, all of which have measured expansion velocities of $\ge$ 2700 km/s \citep{Besw06,Fenech08,Fenech10}, suggesting ages of $\sim$ few hundred years. This can be added to the evidence provided by measured expansion velocities for many of these sources which contradicts the WDB hypothesis of \cite{Sea07}.
The possibility that these compact sources are SNR in the transitional phase from circumstellar to interstellar medium fits well with a similarly observed population of sources within Arp 220, taking into account the higher pressures involved in the more intense star-formation environment within this galaxy. This would also suggest that the more diffuse non-thermal sources within M82 should show little variation in flux density as they would now be expanding into the roughly constant density ISM.
This does not rule out the potential for some of the non-thermal compact sources to be wind-driven bubbles and not SNR. It is interesting to note, however, that the two sources showing flux density variability with measured expansion velocities, pass the feasibility tests outlined in \cite{Sea07}, clearly suggesting that the proposed tests do not adequately distinguish between the WDB and SNR models.

Continued regular monitoring of these sources will give us more insights to their behaviour. It is now possible to carry this out with new sensitive instruments such as {\it e}-MERLIN and the EVLA, and we expect to get a more complete view of the evolution of SNRs in M82. This will allow the investigation of fainter and/or transient objects in starforming galaxies such as M82.

\section{Summary and Conclusion}

We presented the results of the 2009-2010 monitoring sessions of the galaxy M82, obtained with MERLIN and {\it e}-MERLIN at 5 and 6\,GHz respectively. The data include eight epochs between May 2009 and December 2010, and were compared with data from the April 2002 monitoring session of \cite{Fenech08}. 

A total of 52 sources were detected, including three sources absent from the 2002 data: supernova SN2008iz, the transient source 43.78+59.3 and a new SNR shell, 44.28+61.1. Although the first two were previously observed \citep{Brun09,Mux10}, the latter was first detected here.\\

Light curves for all sources have been compiled, including flux density measurements from 2002 \citep{Fenech08}, 2005 \citep{Argo} and these new 2009-2010 data from 5\,GHz MERLIN monitoring sessions. We have found that, in most of the sample, flux densities of SNRs and \ion{H}{ii} regions in M82 stay relatively constant, in agreement with previous analysis \citep[e.g.][]{Kro92,Kro00,Wills98}. Aside from SN2008iz and the well-known variable source 41.95+57.5, two sources show signs of changes in flux density in the short and medium term. These source are among the most compact SNR in M82, possibly showing that flux density variations are due to changes in the circumstellar and interstellar medium in which the shocks travel. They could thus be transitioning from supernovae to supernovae remnants, in analogy with recent results from \cite{Bat11} in Arp220.\\

\section*{Acknowledgements}

MERLIN/e-MERLIN is a national facility operated by The University of Manchester on behalf of the Science and Technology Facilities Council (STFC).


\onecolumn
\appendix
\begin{table}
\section{Flux density measurments}\label{FEpoch}
  \begin{center}
    \caption[]{\label{TotTab}{\sc Normalised} flux-density (as described in \S\ref{FMeas}) for the 2002, 2005 and 2009-2010, MERLIN/{\it e}-MERLIN monitoring sessions of M82. Peak and integrated flux densities are given in mJy/beam and mJy respectively. Spectral index information were retrieved from \cite{McD02}. When unavailable, standard values of $\alpha=-0.7$ for SNR and $\alpha=0.0$ for \ion{H}{ii} regions were assumed. The 2010 6-GHz {\it e}-MERLIN data were converted to 5-GHz using the relation $S_{\nu} \propto \nu^{\alpha}$.}
    \medskip
    \begin{tabular}{rrrrrrrrrr}
      \hline
      \multicolumn{1}{c}{\#} & \multicolumn{1}{c}{Name} & \multicolumn{2}{c}{2009 combined} & \multicolumn{2}{c}{2009-05-01} & \multicolumn{2}{c}{2009-05-26} & \multicolumn{2}{c}{2009-07-28} \\ 
      && \multicolumn{1}{c}{$S_{peak}$} & \multicolumn{1}{c}{$S_{int}$} & \multicolumn{1}{c}{$S_{peak}$} & \multicolumn{1}{c}{$S_{int}$} & \multicolumn{1}{c}{$S_{peak}$} & \multicolumn{1}{c}{$S_{int}$} & \multicolumn{1}{c}{$S_{peak}$} & \multicolumn{1}{c}{$S_{int}$} \\
      \hline
1 & 39.10+57.3 &  0.54 &  4.50$\pm$0.65 &  0.43 &  3.89$\pm$1.20 &  0.71 &  2.72$\pm$2.97 &  0.54 &  3.29$\pm$0.75\\
2 & 39.28+54.1 &  0.23 &  0.23$\pm$0.05 &  0.30 &  0.39$\pm$0.09 &  0.43 &  0.27$\pm$0.21 &  0.26 &  0.44$\pm$0.05\\
3 & 39.40+56.2 &  0.23 &  1.13$\pm$0.29 &  0.24 &  1.29$\pm$0.53 &  0.47 &  1.01$\pm$1.31 &  0.21 &  1.18$\pm$0.33\\
4 & 39.64+53.3 &  0.16 &  0.32$\pm$0.09 &  0.10 &  0.12$\pm$0.17 &       &                &  0.10 &  0.13$\pm$0.10\\
5 & 39.67+55.5 &  0.36 &  1.22$\pm$0.22 &  0.29 &  1.16$\pm$0.40 &  0.39 &  0.66$\pm$0.99 &  0.39 &  1.34$\pm$0.25\\
6 & 39.77+56.9 &  0.16 &  0.36$\pm$0.10 &  0.22 &  0.25$\pm$0.18 &       &                &  0.18 &  0.36$\pm$0.11\\
7 & 40.32+55.2 &  0.28 &  1.35$\pm$0.27 &  0.23 &  0.99$\pm$0.51 &  0.39 &  1.03$\pm$1.06 &  0.24 &  0.95$\pm$0.32\\
8 & 40.61+56.3 &  0.19 &  1.27$\pm$0.33 &  0.24 &  1.34$\pm$0.61 &  0.29 &  0.57$\pm$1.62 &  0.21 &  0.90$\pm$0.38\\
9 & 40.68+55.1 &  0.53 &  7.19$\pm$0.90 &  0.49 &  5.96$\pm$1.67 &  0.70 &  4.85$\pm$4.10 &  0.70 &  5.86$\pm$1.04\\
10& 40.94+58.8 &  0.26 &  0.73$\pm$0.17 &  0.20 &  0.60$\pm$0.32 &       &                &  0.20 &  0.48$\pm$0.19\\
11& 41.18+56.2 &  0.24 &  1.76$\pm$0.44 &  0.26 &  1.43$\pm$0.81 &  0.31 &  0.66$\pm$1.23 &  0.27 &  1.52$\pm$0.50\\
12& 41.30+59.6 &  0.64 &  2.54$\pm$0.29 &  0.57 &  1.89$\pm$0.54 &  0.57 &  2.25$\pm$1.32 &  0.53 &  1.69$\pm$0.34\\
13& 41.64+57.9 &  0.23 &  1.02$\pm$0.22 &  0.23 &  1.24$\pm$0.41 &  0.42 &  0.87$\pm$1.02 &  0.26 &  1.22$\pm$0.26\\
14& 41.95+57.5 & 12.52 & 14.17$\pm$0.19 & 10.80 & 12.63$\pm$0.35 & 12.71 & 14.41$\pm$0.87 & 10.95 & 12.34$\pm$0.22\\
15& 42.08+58.4 &  0.16 &  0.34$\pm$0.09 &  0.34 &  0.48$\pm$0.17 &  0.33 &  0.48$\pm$0.43 &  0.16 &  0.24$\pm$0.11\\
16& 42.20+59.1 &  0.34 &  0.99$\pm$0.19 &  0.36 &  0.50$\pm$0.35 &  0.41 &  0.53$\pm$0.87 &  0.21 &  0.35$\pm$0.22\\
17& 42.43+59.5 &  0.17 &  0.38$\pm$0.15 &  0.31 &  0.50$\pm$0.29 &       &                &  0.15 &  0.40$\pm$0.18\\
18& 42.48+58.4 &  0.32 &  1.37$\pm$0.37 &  0.26 &  1.48$\pm$0.69 &  0.37 &  0.87$\pm$1.69 &  0.22 &  1.25$\pm$0.43\\
19& 42.61+60.7 &  0.20 &  0.67$\pm$0.16 &  0.19 &  0.58$\pm$0.30 &  0.56 &  0.64$\pm$0.74 &  0.20 &  0.52$\pm$0.19\\
20& 42.67+55.6 &  0.21 &  0.81$\pm$0.19 &  0.17 &  0.67$\pm$0.35 &       &                &  0.25 &  0.76$\pm$0.22\\
21& 42.67+56.3 &  0.25 &  0.63$\pm$0.13 &  0.22 &  0.57$\pm$0.24 &  0.44 &  0.83$\pm$0.58 &  0.25 &  0.61$\pm$0.15\\
22& 42.69+58.2 &  0.15 &  0.21$\pm$0.06 &  0.13 &  0.13$\pm$0.11 &       &                &  0.06 &  0.06$\pm$0.07\\
23& 42.80+61.2 &  0.25 &  0.55$\pm$0.11 &  0.26 &  0.52$\pm$0.20 &       &                &  0.20 &  0.46$\pm$0.12\\
24& 42.81+59.6 & 33.18 & 33.57$\pm$0.17 & 30.99 & 32.48$\pm$0.33 & 40.87 & 40.23$\pm$0.80 & 27.24 & 28.34$\pm$0.20\\
25& 43.18+58.2 &  1.30 &  4.79$\pm$0.33 &  0.91 &  4.29$\pm$0.61 &  1.07 &  4.40$\pm$1.50 &  1.28 &  3.77$\pm$0.38\\
26& 43.31+59.2 &  6.93 & 12.52$\pm$0.22 &  5.99 & 10.85$\pm$0.41 &  5.34 &  9.41$\pm$2.97 &  5.77 & 10.52$\pm$0.26\\
27& 43.40+62.6 &  0.19 &  0.61$\pm$0.16 &  0.19 &  0.69$\pm$0.29 &  0.38 &  0.60$\pm$0.71 &  0.19 &  0.48$\pm$0.18\\
28& 43.72+62.6 &  0.26 &  0.42$\pm$0.09 &  0.25 &  0.25$\pm$0.16 &  0.22 &  0.34$\pm$0.39 &  0.16 &  0.19$\pm$0.10\\
29& 43.78+59.3 &  0.84 &  0.90$\pm$0.09 &  0.62 &  0.81$\pm$0.16 &  0.49 &  0.85$\pm$0.39 &  0.59 &  0.64$\pm$0.10\\
30& 43.82+62.8 &  0.16 &  0.33$\pm$0.09 &  0.15 &  0.17$\pm$0.16 &       &                &  0.12 &  0.26$\pm$0.10\\
31& 44.01+59.6 & 13.73 & 25.88$\pm$0.25 & 11.12 & 21.50$\pm$0.46 & 13.52 & 24.84$\pm$1.14 & 11.79 & 22.65$\pm$0.29\\
32& 44.08+63.1 &  0.13 &  0.15$\pm$0.06 &  0.10 &  0.14$\pm$0.11 &       &                &  0.10 &  0.13$\pm$0.07\\
33& 44.28+59.3 &  0.36 &  1.72$\pm$0.32 &  0.48 &  1.60$\pm$0.59 &  0.42 &  1.88$\pm$1.45 &  0.36 &  1.16$\pm$0.37\\
34& 44.28+61.1 &  0.18 &  0.40$\pm$0.11 &  0.21 &  0.40$\pm$0.20 &       &                &  0.17 &  0.23$\pm$0.12\\
35& 44.34+57.8 &  0.22 &  0.98$\pm$0.23 &  0.32 &  1.27$\pm$0.42 &  0.49 &  0.76$\pm$1.03 &  0.21 &  1.03$\pm$0.26\\
36& 44.40+61.8 &  0.26 &  1.19$\pm$0.25 &  0.28 &  0.88$\pm$0.46 &  0.52 &  1.00$\pm$1.12 &  0.19 &  0.84$\pm$0.28\\
37& 44.51+58.2 &  0.26 &  2.38$\pm$0.54 &  0.31 &  1.57$\pm$1.00 &  0.40 &  0.69$\pm$2.46 &  0.26 &  1.51$\pm$0.62\\
38& 44.89+61.2 &  0.32 &  1.16$\pm$0.21 &  0.26 &  0.69$\pm$0.39 &  0.42 &  0.80$\pm$0.95 &  0.35 &  0.90$\pm$0.24\\
39& 44.93+64.0 &  0.20 &  0.39$\pm$0.10 &  0.25 &  0.29$\pm$0.18 &       &                &  0.22 &  0.38$\pm$0.12\\
40& 45.17+61.2 &  3.27 &  8.72$\pm$0.44 &  2.46 &  7.05$\pm$0.81 &  2.78 &  8.12$\pm$2.00 &  2.91 &  6.81$\pm$0.51\\
41& 45.24+65.2 &  0.23 &  1.15$\pm$0.26 &  0.30 &  0.98$\pm$0.49 &  0.82 &  1.63$\pm$1.19 &  0.30 &  0.83$\pm$0.30\\
42& 45.39+60.3 &  0.18 &  1.00$\pm$0.26 &  0.19 &  0.82$\pm$0.49 &  0.37 &  0.66$\pm$1.19 &  0.15 &  0.58$\pm$0.30\\
43& 45.42+67.4 &  0.30 &  2.56$\pm$0.55 &  0.30 &  1.60$\pm$1.02 &  0.41 &  2.25$\pm$2.53 &  0.29 &  1.22$\pm$0.64\\
44& 45.52+64.7 &  0.09 &  0.12$\pm$0.05 &       &                &       &                &  0.11 &  0.13$\pm$0.06\\
45& 45.62+67.0 &  0.26 &  1.54$\pm$0.37 &  0.27 &  1.41$\pm$0.68 &  0.51 &  1.04$\pm$1.68 &  0.17 &  1.17$\pm$0.42\\
46& 45.75+65.3 &  0.21 &  2.24$\pm$0.54 &  0.25 &  1.91$\pm$1.01 &  0.55 &  2.30$\pm$2.48 &  0.25 &  1.94$\pm$0.62\\
47& 45.79+64.0 &  0.22 &  0.49$\pm$0.11 &  0.21 &  0.31$\pm$0.21 &       &                &  0.21 &  0.41$\pm$0.13\\
48& 45.89+63.8 &  0.61 &  1.60$\pm$0.20 &  0.48 &  1.23$\pm$0.38 &  0.48 &  0.80$\pm$0.93 &  0.50 &  1.19$\pm$0.24\\
49& 46.17+67.6 &  0.25 &  2.35$\pm$0.57 &  0.29 &  1.65$\pm$1.05 &  0.60 &  1.54$\pm$2.59 &  0.29 &  1.07$\pm$0.66\\
50& 46.34+66.2 &  0.19 &  0.67$\pm$0.17 &  0.28 &  0.67$\pm$0.32 &       &                &  0.18 &  0.67$\pm$0.19\\
51& 46.52+63.9 &  0.44 &  2.68$\pm$0.47 &  0.42 &  1.96$\pm$0.88 &  0.43 &  1.87$\pm$2.17 &  0.42 &  1.51$\pm$0.55\\
52& 46.70+67.1 &  0.23 &  2.29$\pm$0.54 &  0.27 &  1.29$\pm$0.99 &  0.71 &  1.32$\pm$2.25 &  0.23 &  0.64$\pm$0.62\\
\hline
    \end{tabular}
  \end{center}
\end{table}

\begin{table}
    \centerline{Table~\ref{TotTab} continued.}
  \begin{center}
    \medskip
    \begin{tabular}{rrrrrrrrr}
      \hline
      \multicolumn{1}{c}{\#} & \multicolumn{2}{c}{2009-08-19} & \multicolumn{2}{c}{2009-09-12} & \multicolumn{2}{c}{2009-09-25} & \multicolumn{2}{c}{2010-04-06} \\ 
      & \multicolumn{1}{c}{$S_{peak}$} & \multicolumn{1}{c}{$S_{int}$} & \multicolumn{1}{c}{$S_{peak}$} & \multicolumn{1}{c}{$S_{int}$} & \multicolumn{1}{c}{$S_{peak}$} & \multicolumn{1}{c}{$S_{int}$} & \multicolumn{1}{c}{$S_{peak}$} & \multicolumn{1}{c}{$S_{int}$} \\
      \hline
1 &  0.52 &  4.25$\pm$1.80 &  0.63 &  4.23$\pm$1.38 &  0.54 &  3.77$\pm$1.20 &  0.45 &  3.71$\pm$1.98 \\
2 &  0.34 &  0.39$\pm$0.13 &  0.21 &  0.36$\pm$0.10 &  0.19 &  0.28$\pm$0.09 &  0.21 &  0.22$\pm$0.14 \\
3 &  0.24 &  1.21$\pm$0.80 &  0.18 &  1.09$\pm$0.61 &  0.23 &  0.62$\pm$0.53 &  0.39 &  0.54$\pm$0.33 \\
4 &  0.27 &  0.33$\pm$0.25 &  0.21 &  0.23$\pm$0.20 &  0.11 &  0.14$\pm$0.17 &  0.22 &  0.27$\pm$0.28 \\
5 &  0.32 &  1.16$\pm$0.60 &  0.34 &  1.15$\pm$0.46 &  0.42 &  1.23$\pm$0.40 &  0.33 &  0.62     0.49 \\
6 &  0.27 &  0.35$\pm$0.27 &  0.21 &  0.30$\pm$0.21 &  0.23 &  0.36$\pm$0.18 &  0.40 &  0.48$\pm$0.29 \\
7 &  0.34 &  0.76$\pm$0.76 &  0.46 &  0.96$\pm$0.58 &  0.29 &  0.91$\pm$0.51 &  0.36 &  1.02$\pm$0.83 \\
8 &  0.35 &  1.13$\pm$0.91 &  0.22 &  0.76$\pm$0.70 &  0.34 &  0.82$\pm$0.61 &  0.32 &  0.66$\pm$0.34 \\
9 &  0.80 &  5.86$\pm$2.49 &  0.67 &  6.18$\pm$1.91 &  0.76 &  4.90$\pm$1.67 &  0.56 &  6.02$\pm$2.73 \\
10&  0.33 &  0.75$\pm$0.47 &  0.27 &  0.59$\pm$0.36 &  0.31 &  0.57$\pm$0.32 &       &                \\
11&  0.36 &  1.78$\pm$1.20 &  0.35 &  2.10$\pm$0.93 &  0.22 &  1.12$\pm$0.81 &  0.35 &  0.85$\pm$1.33 \\
12&  0.51 &  1.58$\pm$0.80 &  0.79 &  2.18$\pm$0.62 &  0.43 &  1.66$\pm$0.54 &  0.32 &  0.83$\pm$0.37 \\
13&  0.37 &  1.70$\pm$0.62 &  0.29 &  1.33$\pm$0.48 &  0.31 &  0.70$\pm$0.41 &  0.42 &  0.70$\pm$0.68 \\
14& 11.46 & 12.85$\pm$0.53 & 10.44 & 11.75$\pm$0.41 & 11.24 & 12.63$\pm$0.35 & 10.98 & 12.99$\pm$0.58 \\
15&  0.25 &  0.30$\pm$0.26 &  0.18 &  0.26$\pm$0.20 &  0.17 &  0.18$\pm$0.17 &  0.26 &  0.32$\pm$0.29 \\
16&  0.45 &  0.57$\pm$0.53 &  0.49 &  0.65$\pm$0.41 &  0.20 &  0.63$\pm$0.35 &  0.29 &  0.61$\pm$0.58 \\
17&  0.28 &  0.60$\pm$0.43 &  0.27 &  0.60$\pm$0.33 &  0.25 &  0.47$\pm$0.29 &       &                \\
18&  0.38 &  1.91$\pm$1.03 &  0.42 &  1.67$\pm$0.79 &  0.30 &  1.43$\pm$0.69 &  0.44 &  1.33$\pm$1.13 \\
19&  0.34 &  0.89$\pm$0.45 &  0.29 &  0.81$\pm$0.34 &  0.16 &  0.48$\pm$0.30 &  0.36 &  0.40$\pm$0.49 \\
20&  0.44 &  0.93$\pm$0.52 &  0.24 &  0.83$\pm$0.40 &  0.31 &  0.48$\pm$0.35 &       &                \\
21&  0.23 &  0.54$\pm$0.36 &  0.33 &  0.62$\pm$0.27 &       &                &  0.52 &  0.70$\pm$0.39 \\
22&  0.15 &  0.16$\pm$0.16 &  0.18 &  0.19$\pm$0.12 &  0.12 &  0.16$\pm$0.11 &       &                \\
23&  0.26 &  0.91$\pm$0.30 &  0.47 &  0.81$\pm$0.23 &  0.18 &  0.38$\pm$0.20 &  0.33 &  0.47$\pm$0.32 \\
24& 28.52 & 28.91$\pm$0.49 & 24.52 & 25.38$\pm$0.37 & 26.40 & 26.98$\pm$0.33 & 19.56 & 21.60$\pm$0.53 \\
25&  1.21 &  4.31$\pm$0.91 &  1.02 &  3.98$\pm$0.70 &  0.89 &  3.60$\pm$0.61 &  0.81 &  2.70$\pm$1.00 \\
26&  5.76 & 10.14$\pm$0.61 &  6.01 & 10.93$\pm$0.47 &  5.95 & 10.53$\pm$0.41 &  5.83 & 10.27$\pm$0.68 \\
27&  0.24 &  0.80$\pm$0.44 &  0.21 &  0.67$\pm$0.33 &  0.24 &  0.53$\pm$0.29 &  0.37 &  0.69$\pm$0.47 \\
28&  0.36 &  0.38$\pm$0.24 &  0.15 &  0.24$\pm$0.18 &  0.22 &  0.23$\pm$0.16 &  0.22 &  0.29$\pm$0.26 \\
29&  0.59 &  0.58$\pm$0.24 &  0.53 &  0.50$\pm$0.18 &  0.80 &  0.72$\pm$0.16 &  0.56 &  0.60$\pm$0.26 \\
30&  0.35 &  0.54$\pm$0.25 &  0.24 &  0.35$\pm$0.19 &  0.20 &  0.25$\pm$0.16 &       &                \\
31& 12.01 & 22.52$\pm$0.69 & 11.58 & 21.90$\pm$0.53 & 12.50 & 23.48$\pm$0.46 & 11.95 & 24.16$\pm$0.76 \\
32&       &                &       &                &  0.09 &  0.11$\pm$0.11 &       &                \\
33&  0.50 &  1.54$\pm$0.88 &  0.39 &  1.25$\pm$0.68 &  0.56 &  1.69$\pm$0.59 &  0.28 &  1.03$\pm$0.97 \\
34&       &                &       &                &       &                &       &                \\
35&  0.35 &  1.37$\pm$0.63 &  0.36 &  1.24$\pm$0.48 &  0.21 &  0.48$\pm$0.56 &  0.37 &  1.01$\pm$0.69 \\
36&  0.25 &  0.93$\pm$0.68 &  0.27 &  0.75$\pm$0.52 &  0.25 &  0.79$\pm$0.46 &  0.29 &  0.96$\pm$0.75 \\
37&  0.38 &  1.35$\pm$1.49 &  0.32 &  1.86$\pm$1.14 &  0.35 &  1.04$\pm$1.00 &  0.41 &  0.63$\pm$1.64 \\
38&  0.43 &  0.67$\pm$0.58 &  0.39 &  0.81$\pm$0.44 &  0.27 &  0.81$\pm$0.39 &  0.48 &  1.02$\pm$0.63 \\
39&  0.23 &  0.24$\pm$0.27 &  0.42 &  0.46$\pm$0.21 &       &                &  0.19 &  0.22$\pm$0.30 \\
40&  3.08 &  7.09$\pm$1.21 &  2.72 &  6.39$\pm$0.94 &  2.95 &  7.64$\pm$0.81 &  2.40 &  6.32$\pm$1.33 \\
41&  0.46 &  1.57$\pm$0.72 &  0.36 &  1.23$\pm$0.55 &  0.27 &  1.00$\pm$0.49 &  0.37 &  0.71$\pm$0.80 \\
42&  0.38 &  0.99$\pm$0.72 &  0.31 &  0.82$\pm$0.55 &  0.36 &  0.77$\pm$0.49 &  0.36 &  0.99$\pm$0.28 \\
43&  0.50 &  1.04$\pm$1.53 &  0.33 &  1.74$\pm$1.18 &  0.35 &  1.47$\pm$1.02 &  0.48 &  1.57$\pm$1.68 \\
44&       &                &       &                &       &                &       &                \\
45&  0.43 &  2.09$\pm$1.02 &  0.29 &  1.42$\pm$0.78 &  0.23 &  0.94$\pm$0.68 &  0.23 &  0.54$\pm$1.12 \\
46&  0.51 &  3.07$\pm$1.50 &  0.33 &  2.80$\pm$1.16 &  0.35 &  2.09$\pm$1.01 &  0.35 &  1.80$\pm$1.65 \\
47&  0.37 &  0.58$\pm$0.31 &  0.26 &  0.57$\pm$0.24 &  0.28 &  0.59$\pm$0.21 &  0.25 &  0.35$\pm$0.34 \\
48&  0.41 &  0.96$\pm$0.57 &  0.55 &  1.20$\pm$0.43 &  0.41 &  0.98$\pm$0.38 &  0.31 &  1.07$\pm$0.62 \\
49&  0.33 &  1.89$\pm$1.57 &  0.34 &  0.67$\pm$1.21 &  0.27 &  0.81$\pm$1.05 &  0.47 &  1.68$\pm$1.72 \\
50&  0.38 &  1.16$\pm$0.47 &  0.21 &  0.56$\pm$0.36 &  0.21 &  0.39$\pm$0.32 &       &                \\
51&  0.61 &  2.05$\pm$1.32 &  0.37 &  1.96$\pm$1.01 &  0.44 &  2.41$\pm$0.88 &  0.47 &  1.44$\pm$1.44 \\
52&  0.36 &  1.86$\pm$1.48 &  0.34 &  2.25$\pm$1.14 &  0.36 &  1.34$\pm$0.99 &  0.35 &  0.48$\pm$1.63 \\
\hline
    \end{tabular}
  \end{center}
\end{table}

\begin{table}
  \begin{center}
    \centerline{Table~\ref{TotTab} continued.}
    \medskip
    \begin{tabular}{rrrrrr}
      \hline
      \multicolumn{1}{c}{\#} & \multicolumn{2}{c}{2010-12-17} & \multicolumn{1}{c}{2002} & \multicolumn{1}{c}{2005} &\multicolumn{1}{c}{$\alpha$} \\ 
      &\multicolumn{1}{c}{$S_{peak}$} & \multicolumn{1}{c}{$S_{int}$} & \multicolumn{1}{c}{$S_{int}$} & \multicolumn{1}{c}{$S_{int}$} &\\
      \hline
1 &  0.52 &  4.22$\pm$0.91 &  3.85$\pm$0.31 &  3.00$\pm$0.77 &$-$0.53 \\
2 &  0.21 &  0.41$\pm$0.11 &  0.51$\pm$0.07 &  0.24$\pm$0.17 &   1.64 \\
3 &  0.25 &  0.63$\pm$0.43 &  0.89$\pm$0.22 &  0.49$\pm$0.16 &$-$1.04 \\
4 &  0.20 &  0.27$\pm$0.17 &  0.11$\pm$0.02 &  0.13$\pm$0.05 &$-$0.71 \\
5 &  0.40 &  0.80$\pm$0.18 &  0.87$\pm$0.08 &  0.57$\pm$0.17 &   1.03 \\
6 &  0.18 &  0.23$\pm$0.07 &  0.19$\pm$0.02 &                &        \\
7 &  0.27 &  0.60$\pm$0.16 &  0.27$\pm$0.04 &                &$-$0.23 \\
8 &  0.25 &  0.78$\pm$0.33 &  0.44$\pm$0.07 &                &$-$0.72 \\
9 &  0.56 &  6.06$\pm$0.99 &  6.60$\pm$0.28 &  5.34$\pm$1.06 &$-$0.54 \\
10&  0.19 &  0.48$\pm$0.16 &  0.29$\pm$0.07 &                &   0.44 \\
11&  0.20 &  1.27$\pm$0.41 &  0.90$\pm$0.14 &  0.85$\pm$0.28 &   0.87 \\
12&  0.48 &  2.04$\pm$0.38 &  2.49$\pm$0.28 &  1.88$\pm$0.32 &$-$0.47 \\
13&  0.17 &  0.67$\pm$0.19 &  0.26$\pm$0.05 &  0.45$\pm$0.13 &   1.32 \\
14& 10.04 & 12.34$\pm$0.32 & 18.45$\pm$0.23 & 17.25$\pm$0.30 &$-$0.80 \\
15&  0.21 &  0.40$\pm$0.14 &  0.35$\pm$0.06 &                &   1.32 \\
16&  0.25 &  0.89$\pm$0.29 &  1.12$\pm$0.16 &                &   1.16 \\
17&       &                &  0.19$\pm$0.02 &                &        \\
18&  0.17 &  0.89$\pm$0.32 &  1.05$\pm$0.19 &                &$>$1.20 \\
19&  0.24 &  0.37$\pm$0.17 &  0.32$\pm$0.07 &                &$-$0.70 \\
20&  0.30 &  0.57$\pm$0.30 &  1.20$\pm$0.29 &                &$-$1.30 \\
21&       &                &  0.91$\pm$0.17 &                &        \\
22&  0.16 &  0.33$\pm$0.18 &  0.59$\pm$0.14 &  0.48$\pm$0.12 &   1.04 \\
23&  0.22 &  0.44$\pm$0.12 &  0.59$\pm$0.08 &  0.26$\pm$0.10 &$-$0.63 \\
24& 13.35 & 14.77$\pm$0.27 &                &                &        \\
25&  0.80 &  2.98$\pm$0.33 &  4.23$\pm$0.34 &  2.93$\pm$0.37 &$-$0.44 \\
26&  5.69 & 10.12$\pm$0.29 & 10.32$\pm$0.11 & 10.81$\pm$0.33 &$-$0.65 \\
27&  0.18 &  0.23$\pm$0.08 &  0.49$\pm$0.15 &                &$-$0.70 \\
28&  0.18 &  0.20$\pm$0.09 &  0.30$\pm$0.04 &  0.18$\pm$0.07 &        \\
29&  0.86 &  1.02$\pm$0.10 &                &                &$-$0.70 \\
30&  0.21 &  0.34$\pm$0.14 &  0.35$\pm$0.05 &  0.64$\pm$0.17 &$-$0.70 \\
31& 12.40 & 23.74$\pm$0.39 & 21.04$\pm$0.22 & 25.25$\pm$0.42 &$-$0.38 \\
32&       &                &  0.43$\pm$0.07 &                &        \\
33&  0.20 &  0.96$\pm$0.33 &  1.78$\pm$0.17 &  0.31$\pm$0.12 &$-$0.72 \\
34&  0.14 &  0.19$\pm$0.08 &                &                &        \\
35&  0.28 &  0.80$\pm$0.29 &  1.03$\pm$0.16 &                &$-$0.70 \\
36&  0.26 &  0.97$\pm$0.31 &  0.68$\pm$0.17 &                &   0.07 \\
37&  0.23 &  0.99$\pm$0.35 &  1.57$\pm$0.19 &  0.53$\pm$0.18 &$-$0.15 \\
38&  0.23 &  0.81$\pm$0.28 &  0.93$\pm$0.10 &                &$-$0.45 \\
39&       &                &  0.23$\pm$0.05 &                &   1.10 \\
40&  2.61 &  6.49$\pm$0.50 &  6.83$\pm$0.21 &  6.53$\pm$0.45 &$-$0.52 \\
41&  0.33 &  0.66$\pm$0.16 &  1.15$\pm$0.17 &                &$-$1.02 \\
42&       &                &  0.48$\pm$0.13 &                &        \\
43&  0.37 &  1.30$\pm$0.40 &  1.25$\pm$0.15 &  0.82$\pm$0.26 &$-$0.83 \\
44&  0.14 &  0.20$\pm$0.11 &  0.19$\pm$0.04 &                &        \\
45&  0.19 &  0.73$\pm$0.30 &  0.92$\pm$0.16 &                &        \\
46&  0.26 &  1.64$\pm$0.57 &  1.65$\pm$0.33 &                &$-$0.70 \\
47&  0.18 &  0.29$\pm$0.12 &  0.21$\pm$0.03 &                &$-$0.55 \\
48&  0.42 &  0.87$\pm$0.25 &  1.34$\pm$0.08 &  0.64$\pm$0.19 &$-$0.38 \\
49&  0.18 &  1.08$\pm$0.41 &  1.24$\pm$0.28 &                &   0.66 \\
50&       &                &  0.22$\pm$0.03 &                &        \\
51&  0.32 &  1.67$\pm$0.48 &  0.70$\pm$0.20 &                &$-$0.20 \\
52&  0.24 &  1.18$\pm$0.46 &  1.56$\pm$0.41 &  0.77$\pm$0.24 &$-$0.80 \\
\hline
    \end{tabular}
  \end{center}
\end{table}

\label{lastpage}

\end{document}